\documentclass[conference,a4paper,10pt]{IEEEtran}
\IEEEoverridecommandlockouts

\usepackage[utf8]{inputenc}
\usepackage[T1]{fontenc}
\usepackage[a4paper, total={17cm,25cm}]{geometry}
\usepackage{url}
\usepackage{amsmath,amssymb}
\usepackage{amsthm}
\usepackage{enumerate}
\usepackage{booktabs}
\usepackage{tabularx}
\usepackage{pgfplots}
\usetikzlibrary{positioning}
\usetikzlibrary{shapes.misc,shapes.geometric}
\pgfplotsset{compat=1.18}
\usepackage[ruled,vlined]{algorithm2e}
\usepackage[capitalize]{cleveref}
\usepackage{enumerate}
\usepackage[normalem]{ulem}
\usepackage{bbm}
\usepackage{balance}
\usepackage{subfigure}
\usepackage{cite}
\usepackage{comment}
\usepackage{cancel}

\usepackage[nolist,nohyperlinks]{acronym}
\acrodef{ldp}[LDP]{local differential privacy}
\acrodef{3r}[3R]{Three-Region}
\acrodef{mq}[MQ]{Markov-Quilt}

\newtheorem{definition}{Definition}
\newtheorem{remark}{Remark}
\newtheorem{example}{Example}
\newtheorem{lemma}{Lemma}
\newtheorem{proposition}{Proposition}
\newtheorem{corollary}{Corollary}
\newtheorem{theorem}{Theorem}

\definecolor{asparagus}{rgb}{0.53, 0.66, 0.42}
\definecolor{bostonuniversityred}{rgb}{0.8, 0.0, 0.0}
\definecolor{mq}{RGB}{46,37,133}
\definecolor{3rConvex}{RGB}{51,117,56}
\definecolor{3rNumerical}{RGB}{194,106,119}

\DeclareMathOperator{\supp}{supp}

\newcommand{\M}{\mathcal{M}}
\newcommand{\X}{\mathbf{X}}
\newcommand{\x}{\mathbf{x}}
\newcommand{\Y}{\mathbf{Y}}
\newcommand{\y}{\mathbf{y}}

\renewcommand{\P}{\mathcal{P}}
\newcommand{\p}{p}
\newcommand{\R}{\mathcal{R}}

\newcommand{\Q}{\mathcal{Q}}
\renewcommand{\Pr}[1]{{\operatorname{Pr}\left( {#1} \right)}}
\newcommand{\range}[1]{{\left[ {#1} \right]}}

\newcommand{\indicator}[1]{{\mathbbm{1}\left( {#1} \right)}}

\newcommand{\Exp}[2][]{{\ensuremath{\mathbb{E}_{#1}\left[{#2}\right]}}}

\newcommand{\leak}[1]{{\mathcal{L}(#1)}}
\newcommand{\infl}[1]{{\mathcal{I}(#1)}}
\newcommand{\pwInfl}[1]{{i(#1)}}
\newcommand{\rr}[1][]{\rho_{#1}}

\newcommand{\highInfl}[1]{{\hat{i}\left({#1}\right)\,\xspace}}
\newcommand{\lowInfl}[1]{{\check{i}\left({#1}\right)\,\xspace}}

\newcommand{\region}[1][]{\ensuremath{{\mathcal{R}_{\mathrm{#1}}}}}
\newcommand{\recordS}{\mathcal{S}}
\newcommand{\recordM}{\mathcal{M}}
\newcommand{\recordL}{\mathcal{L}}

\newcommand{\define}{\ensuremath{\triangleq}}

\newcommand{\dd}{data-dependent\xspace}
\newcommand{\di}{data-independent\xspace}

\newcommand{\one}{\boldsymbol{1}}
\newcommand{\zero}{\boldsymbol{0}}
\newcommand{\powerset}{\mathcal{P}}

\begin{document}

\title{Between Close Enough to Reveal and Far Enough to Protect: a New Privacy Region for Correlated Data}

\author{
    \vspace{-1cm}
	\IEEEauthorblockN{Luis Maßny, Rawad Bitar, Fangwei Ye, Salim El Rouayheb}
    \thanks{LM and RB are with the School of Computation, Information and Technology at the Technical University of Munich, Germany. Emails: \{luis.massny, rawad.bitar\}@tum.de}
    \thanks{FY is with the College of Computer Science and Technology at the Nanjing University of Aeronautics and Astronautics, Nanjing, China. Email: fangweiye@nuaa.edu.cn}
    \thanks{SER is with the ECE Department at Rutgers University, New Brunswick, NJ, USA. Email: salim.elrouayheb@rutgers.edu}
    \thanks{This project is funded by the Bavarian Ministry of Economic Affairs, Regional Development and Energy within the scope of the 6G Future Lab Bavaria, and by DFG (German Research Foundation) projects under Grant Agreement No. BI 2492/1-1.}
}

\maketitle

\begin{abstract}
    When users make personal privacy choices, correlation between their data can cause inadvertent leakage about users who do not want to share their data through other users sharing their data.
    As a solution, we consider local redaction mechanisms.
    To model pre-existing approaches, we study the class of \di privatization mechanisms within this framework and upper-bound their utility when data correlation is modeled by a stationary Markov process. In contrast, we find a novel family of \dd mechanisms, which improve the utility by leveraging a \dd leakage measure.
\end{abstract}

\section{Introduction}

We consider the problem of releasing correlated data records located at one or multiple data owners. However, some of the records must remain private in a differential privacy sense. That is, a privacy-preserving data processing is required, called \emph{mechanism}, where we focus on local redaction (erasure) mechanisms. This is the setting depicted in \cref{fig:mc_redaction_setting}. We ask \emph{how many records can be revealed in this setting while protecting the record of a user requesting privacy}?

A straightforward solution would be to separate the records into one region with low correlation and one region with high correlation, and redact the latter, cf.~\cite{songPufferfishPrivacyMechanisms2017a}.
Surprisingly, we show that one can do better. Towards that end, we focus on the Markov setting and show that there exists a region in which records have high correlation, but must not always be redacted.
The main idea is to leverage \dd leakage information, which allows for a more granular redaction decision.

Our motivation for this problem is the important role of data privacy in modern networked ecosystems, which therefore, is also the subject of extensive laws, such as the European Union's General Data Protection Regulation (GDPR)~\cite{europeanparliamentRegulationEU2016} and the California Consumer Privacy Act (CCPA)~\cite{ccpa2018}. However, data privacy is not yet fully achieved; in particular, when concerning the user's right to opt out from sharing their personal data and the right to be forgotten.
Only erasing the data of the user in question (as required by law) is not enough. This is because many types of datasets contain correlated records, such as time series data and location traces~\cite{liuLocationPrivacyIts2018}, or due to natural correlation in social networks or on social media platforms~\cite{lucaHowSocialRelationships2010,bonhardKnowingMeKnowing2006}.
Despite its erasure, information about a record can be inferred from non-erased correlated records in such a dataset. Moreover, leaking correlated data can expose personal information even about individuals who are not present in the dataset~\cite{rebeccacarballoDataBreach23andMe2023}.
Our focus on redaction mechanisms is also motivated by the fact that perturbations might be undesired since they reduce the faithfulness of the data~\cite{yeMechanismsHidingSensitive2022}, and may require individuals to give (socially) undesired responses, e.g., asking a user to report being sick while being healthy or to badly rate a movie they liked.
Perturbation-based mechanisms as an alternative to redaction are left for future work.
Due to our interest in distributed settings, we focus on local mechanisms. As such, each user owning their private record can apply the redaction mechanism independently from other users.
Besides the applicability to distributed settings, local mechanisms are attractive owing to their small algorithmic complexity, memory efficiency, and ability to operate on online data.

\begin{figure}[t]
    \centering
    \begin{tikzpicture}
    \tikzset{
        sample/.style={minimum width=1cm,inner sep=0}
    }
    \node[sample] (x1) at (0,0) {$X_1$};
    \node (dots1) at (1,0) {$\dots$};
    \node[draw,circle,sample,label=above:{\footnotesize private}] (xp) at (2.5,0) {$X_\p$};
    \node (dots2) at (4,0) {$\dots$};
    \node[sample] (xn) at (5,0) {$X_n$};

    \node[draw,rectangle, font=\small] (m1) at (0,-1) {$\Pr{Y_1 | X_1}$};
    \node[draw,rectangle, font=\small] (mp) at (2.5,-1) {$\Pr{Y_\p | X_\p}$};
    \node[draw,rectangle, font=\small] (mn) at (5,-1) {$\Pr{Y_n | X_n}$};
    
    \node (y1) at (0,-2) {$Y_1 = X_1$};
    \node (yp) at (2.5,-2) {$Y_\p = \perp$};
    \node (yn) at (5,-2) {$Y_n = X_n$};

    \draw (x1) edge[-] (dots1);
    \draw (dots1) edge[-] (xp);
    \draw (xp) edge[-] (dots2);
    \draw (dots2) edge [-] (xn);
    
    \draw[->] (x1) edge (m1) (m1) edge (y1);
    \draw[->] (xp) edge (mp) (mp) edge (yp);
    \draw[->] (xn) edge (mn) (mn) edge (yn);

    \draw[dashed] (-1.5,-0.5) rectangle (6.5,-1.5);
    \node[anchor=east] (m) at (-0.9,-1) {$\mathcal{M}$};
\end{tikzpicture}
    \caption{Depiction of the problem setting. The goal is to release as many of the correlated records $X_1,\dots,X_n$ while preserving \acl{ldp} of $X_\p$. A mechanism $\M$ is required, which operates locally on each $X_t \in \{0,1\}$ and outputs $Y_t \in \{X_t,\perp\}$, which can be a redaction ($\perp$), according to a distribution $\Pr{Y_t|X_t}$ for each $t \in \range{n}$.}
    \label{fig:mc_redaction_setting}
\end{figure}

\subsection{Related Work}
The challenge of ensuring data privacy in databases with correlated records has been studied in a rich line of work~\cite{kiferNoFreeLunch2011,chenCorrelatedNetworkData2014,xiaoProtectingLocationsDifferential2015,yangBayesianDifferentialPrivacy2015,zhuCorrelatedDifferentialPrivacy2015,liuDependenceMakesYou2016,ghoshInferentialPrivacyGuarantees2017,songPufferfishPrivacyMechanisms2017a,chakrabartiOptimalLocalBayesian2022}. The popular framework of differential privacy~\cite{dworkDifferentialPrivacy2006} in its original definition provides privacy guarantees that are only meaningful for independent records~\cite{kiferNoFreeLunch2011}. To overcome this limitation, previous works have proposed tailored privacy measures for the case of correlated data.
A popular solution to account for correlations is to employ a differentially private mechanism with stricter privacy parameters. Depending on the correlation model, the privacy parameter is chosen to satisfy a group differential privacy requirement~\cite{chenCorrelatedNetworkData2014}, dependent differential privacy requirement~\cite{zhaoDependentDifferentialPrivacy2017}, or inferential privacy requirement~\cite{ghoshInferentialPrivacyGuarantees2017}. More specifically, the so-called Laplace mechanism is employed with a scale parameter tuned according to the correlation between the data and the privacy requirement, e.g., \cite{liuDependenceMakesYou2016,zhuCorrelatedDifferentialPrivacy2015,songPufferfishPrivacyMechanisms2017a,xiaoProtectingLocationsDifferential2015,yangBayesianDifferentialPrivacy2015}. 
Notably, all of the aforementioned works consider settings in which the database is owned by a single entity. Therefore, \emph{centralized perturbation mechanisms} that operate on the data as a whole can be employed. The work of~\cite{chakrabartiOptimalLocalBayesian2022} considers \emph{local perturbation mechanisms} that perturb the records independently

In a different line of work, \cite{naimONOFFPrivacyCorrelated2019a,yeIntermittentPrivateInformation2022,jiangAnsweringCountQueries2023,yeMechanismsHidingSensitive2022,naimPrivacySocialNetworks2023} consider a granular privacy requirement. Instead of ensuring a privacy guarantee for the whole dataset, only the privacy of a specific subset of the records is considered, which is also the focus of this work.
Namely, the so-called problem of ON-OFF privacy with perfect information-theoretic privacy is studied in the context of private information retrieval~\cite{naimONOFFPrivacyCorrelated2019a,yeIntermittentPrivateInformation2022} and genomic data analysis~\cite{yeMechanismsHidingSensitive2022,jiangAnsweringCountQueries2023}. For dependent differential privacy guarantees in an ON-OFF setting, a perturbation mechanism has been developed in~\cite{naimPrivacySocialNetworks2023}. Similarly to the previous line of work, the privatization mechanisms used here are not local, i.e., they require access to the whole or big parts of the dataset.

\subsection{Contributions}
In this work, we study the ON-OFF privacy problem using local privatization mechanisms, i.e., we require the mechanism to access only one record at a time. We consider a \ac{ldp}~\cite{duchiLocalPrivacyStatistical2013} requirement and focus on \emph{redaction mechanisms} which either release the true record or substitute it with a redaction symbol (erasure). We present our results for binary data records with correlation modeled by a Markov chain.
Our particular contributions are:
\begin{itemize}
    \item we study the limits of pre-existing approaches, namely, \di mechanisms, in the local redaction setting,
    \item we give a novel family of local redaction mechanisms, which leverages \dd leakage information,
    \item we show and numerically demonstrate that the novel mechanisms improve the utility over \di mechanisms. As a first step, we provide a simple mechanism design rule using a convex relaxation.
\end{itemize}

\section{Problem Setting}
\label{sec:setting}

\paragraph*{Notation}
We define $\range{n} \define \{1,\dots,n\}$. Vectors are represented by bold letters, e.g., $\X$ and $\x$, and sets are represented by calligraphic letters, e.g., $\mathcal{Q}$. Random variables and random vectors are denoted by upper-case letters $X$ and $\X$, respectively, and their realizations are denoted by the same lower-case letter, i.e., $x$ and $\x$, respectively.
The power set of a set $\mathcal{Q}$ is denoted by $\powerset(\mathcal{Q})$.
For a random variable $X \in \mathcal{X}$, we define its support as $\supp(X) \define \{ x \in \mathcal{X} \colon \Pr{X=x} > 0 \}$. The support of a random vector is defined accordingly.
Given $n$ random variables $X_1,\dots, X_n$ and a set $\mathcal{Q} \subset \range{n}$, we let $\X_\mathcal{Q}$ be the vector of $X_i$'s indexed by $\mathcal{Q}$, i.e., $\X_\mathcal{Q} \define (X_i)_{i\in\mathcal{Q}}$. The same holds for their realizations, i.e., $\mathbf{x}_\mathcal{Q}\define (x_i)_{i\in\mathcal{Q}}$. For $x \in \{0,1\}$, we define its complement as $\bar{x} \define 1-x$.

\paragraph*{Problem setting}
We consider a setting in which a set of $n$ individuals hold binary data records $X_1,\dots, X_n$, $X_t \in \mathcal{X}\define\{0,1\}$ and an analyst requesting to know their realizations $x_1,\dots, x_n$. The records $X_t$ are identically distributed but dependent. An individual $\p \in [n]$ requires privacy and does not want to reveal the realization of their record. Due to the correlation in the data, other individuals must also not reveal the realization of their records to help guarantee the privacy of $X_\p$. The goal is to design a redaction mechanism that reveals as many records from $\X \define (X_1,\dots,X_n)$ as possible to the data analyst while preserving the privacy of $X_\p$.%
\footnote{Our solutions can be extended to preserve privacy for several $\P \subset \range{n}$ (but with potentially lower performance) by applying it to each $\p \in \P$ separately, and choosing the most conservative redaction model locally.}

We consider \emph{local redaction mechanisms} $\M \colon \mathcal{X}^n \to \mathcal{Y}^n$, which output a privatized data vector $\Y=(Y_1,\dots,Y_n)$. The mechanism either outputs $Y_t = X_t$ or redacts $X_t$ and outputs an erasure $\perp$ instead, i.e., $Y_t \in \{X_t,\perp\}$ and $\mathcal{Y} = \mathcal{X} \cup \{\perp\}$. Since the redaction mechanism is local, the following holds: $\Pr{ Y_t=y_t | \X=\x } = \Pr{ Y_t=y_t | X_t=x_t }$.

To model the correlation between the records, we assume that $X_1,\dots, X_n$ form a Markov chain $X_1 - X_2 - \cdots - X_n$ with transition matrix $P(t)$, where for $i,j \in \{0,1\}$, $P_{ij}(t) = \Pr{ X_{t+1}=j | X_{t}=i }$.
We consider a stationary transition matrix $P(t) = P$ with $P_{01} = \alpha$ and $P_{10} = \beta$ for all $t\in [n]$ with $0 < \alpha \leq \beta < 1$, i.e.,
\begin{equation*}
    P(t)=P =
    \begin{pmatrix}
        1-\alpha & \alpha \\
        \beta & 1-\beta
    \end{pmatrix}.
\end{equation*}
Since the records $X_t$ are identically distributed, this means that the marginal distribution is the Markov chain's stationary distribution, i.e., $\Pr{X_t = 0} = \frac{\beta}{\alpha+\beta}$.
In this case, it holds that $\Pr{ X_{t+1}=j | X_{t}=i } = \Pr{ X_{t}=j | X_{t+1}=i }$. Therefore, the forward transition probabilities are the same as the backward transition probabilities.
The setting is depicted in \cref{fig:mc_redaction_setting}. We say that $X_t$ is left (right, respectively) of $X_\p$, when $t \leq \p$ ($t \geq \p$, respectively).
For a set $\mathcal{Q} \subseteq \range{n}$, we use $\mathcal{Q}^{(\ell)} \define \mathcal{Q} \cap [1,\p]$ and $\mathcal{Q}^{(r)} \define \mathcal{Q} \cap [\p,n]$ to denote the elements left and right of $X_\p$, respectively.
W.l.o.g., we assume that $0 \leq \p \leq n/2$ (if not, we can re-index the elements). 

\paragraph*{Definitions}
As the privacy measure, we adopt \ac{ldp}~\cite{duchiLocalPrivacyStatistical2013,ghoshInferentialPrivacyGuarantees2017} of record $X_\p$ when given the output $\Y$ defined next.

\begin{definition}[LDP]
\label{def:ldp}
A local mechanism $\M \colon \mathcal{X}^n \to \mathcal{Y}^n,$ with input $\X$ and output $\Y$ has a \emph{privacy leakage} $\leak{X_\p \to \Y}$ about $X_\p \in \mathcal{X}$ into $\Y$, defined as
\begin{equation*}
    \leak{X_\p \to \Y} \define \log \sup_{\y \in \supp(\Y), \, x \in \mathcal{X}} \, \frac{ \Pr{ \Y=\y | X_\p=x} }{ \Pr{ \Y=\y | X_\p=\bar{x} } }.
\end{equation*}
The mechanism $\M$ is $\epsilon$-private about $X_\p$ if
    $\leak{X_\p \to \Y} \leq \epsilon$.
\end{definition}
\noindent When the goal is to find an $\epsilon$-private mechanism, we refer to $\epsilon$ as the \emph{privacy budget}.
\ac{ldp} belongs to the class of pufferfish privacy measures~\cite{kiferPufferfishFrameworkMathematical2014}. For local mechanisms, it is stronger than differential privacy~\cite{dworkDifferentialPrivacy2006}, and is closely related to the even stricter notions of dependent and Bayesian differential privacy~\cite{zhaoDependentDifferentialPrivacy2017,yangBayesianDifferentialPrivacy2015}.
For more information on related privacy measures, we refer the reader to the comprehensive surveys~\cite{blochOverviewInformationTheoreticSecurity2021,wagnerTechnicalPrivacyMetrics2018}.

As the utility measure, we adopt the expected fraction of correctly released records~\cite{yeMechanismsHidingSensitive2022}.
\begin{definition}[Utility]
The utility $\nu$ of a redaction mechanism $\M \colon \mathcal{X}^n \to \mathcal{Y}^n$ with input $\X$ and output $\Y$ is
\begin{equation*}
    \nu \define \frac{1}{n} \Exp{ \sum_{t=1}^{n} \indicator{X_t=Y_t} },
\end{equation*}
where $\indicator{\cdot}$ denotes the indicator function.
\end{definition}
\noindent This definition coincides with the Hamming distance~\cite{chakrabartiOptimalLocalBayesian2022,naimPrivacySocialNetworks2023}, $L_1$-distance~\cite{songPufferfishPrivacyMechanisms2017a}, and $L_2$-distance~\cite{xiaoProtectingLocationsDifferential2015} for binary records.

We also introduce a set of terms and symbols that are required to represent our main results. As a dependence measure, we use the so-called influence from a record $X_\p$ on the realization of records $\x_\mathcal{S}$. The largest influence among the values $\x_\mathcal{S}$ is known as the max-influence ~\cite{songPufferfishPrivacyMechanisms2017a,zhaoDependentDifferentialPrivacy2017,naimPrivacySocialNetworks2023}.

\begin{definition}[Pointwise-influence and max-influence]
The \emph{pointwise-influence} from a record $X_\p$ on realizations of the records in $\mathcal{S} \subseteq \range{n} \setminus \{\p\}$, is defined as
\begin{equation*}
\pwInfl{X_\p \rightsquigarrow \X_\mathcal{S}=\x_\mathcal{S}}
\define
\log \max_{x \in \mathcal{X}} \frac{ \Pr{\X_\mathcal{S}=\x_\mathcal{S} | X_\p=x} }{ \Pr{\X_\mathcal{S}=\x_\mathcal{S} | X_\p=\bar{x}} }.
\end{equation*}
The \emph{max-influence} from a record $X_\p$ on records $\X_\mathcal{S}$ is defined as
\begin{equation*}
    \infl{X_\p \rightsquigarrow \X_\mathcal{S}} \define \max_{\x_\mathcal{S} \in \mathcal{X}^{|\mathcal{S}|}} \pwInfl{X_\p \rightsquigarrow \X_\mathcal{S}=\x_\mathcal{S}},
\end{equation*}
where we define $\infl{X_\p \rightsquigarrow \X_\emptyset} = 0$ and $\infl{X_\p \rightsquigarrow X_\p} = \infty$.
\end{definition}

The pointwise-influence $\pwInfl{X_\p \rightsquigarrow X_t=x_t}$ defines a separation of the records into three (possibly empty) sets, called regions, for a parameter $0 < \epsilon^\prime \leq \epsilon$:
\begin{align*}
    \region[L|\epsilon^\prime] &\define \{ t \in \range{n} \colon \epsilon^\prime < \pwInfl{X_\p \rightsquigarrow X_t=0} \leq \pwInfl{X_\p \rightsquigarrow X_t=1} \},\\
    \region[M|\epsilon^\prime] &\define \{ t \in \range{n} \colon \pwInfl{X_\p \rightsquigarrow X_t=0} \leq \epsilon^\prime < \pwInfl{X_\p \rightsquigarrow X_t=1} \},\\
    \region[S|\epsilon^\prime] &\define \{ t \in \range{n} \colon \pwInfl{X_\p \rightsquigarrow X_t=0} \leq \pwInfl{X_\p \rightsquigarrow X_t=1} \leq \epsilon^\prime \}.
\end{align*}

\section{Main Results}
\label{sec:main_results}

\begin{figure}[t]
    \resizebox{\linewidth}{!}{\input{redaction_regions}}
    \caption{Pointwise-influence about $X_\p$ for $\alpha=0.25,\beta=0.5$.
    Records with a pointwise-influence of more than $\epsilon^\prime$ are always redacted (${\region[L|\epsilon^\prime]}$).
    Records with a pointwise-influence of at most $\epsilon^\prime$ can be released potentially (${\region[M|\epsilon^\prime]}$).
    Records with a max-influence of less than $\epsilon^\prime$ can always be released ($\region[S|\epsilon^\prime]$).
    The \ac{3r} mechanism uses $\epsilon^\prime=\epsilon/2$ in this example.}
    \label{fig:redaction_regions}
\end{figure}

Our main result is a novel family of local redaction mechanisms, named \ac{3r} mechanisms.
A \ac{3r} mechanism separates the records into three regions around the private record $X_\p$, as illustrated in \cref{fig:redaction_regions}. In each region, $\region[S|\epsilon^\prime],\region[M|\epsilon^\prime],\region[L|\epsilon^\prime] \subseteq \range{n}$ (for a parameter $\epsilon^\prime$), a \ac{3r} mechanism takes a different redaction approach. Most importantly, it improves over prior approaches by employing a \dd redaction strategy in the region $\region[M|\epsilon^\prime]$.
We give its privacy-utility tradeoff in \cref{thm:randomized_redaction}, which is derived in \cref{sec:3r_mechanism}.

\begin{theorem}
\label{thm:randomized_redaction}
A \ac{3r} mechanism is $\epsilon$-private, for a chosen $\epsilon>0$, and can achieve a utility $\nu_{\mathrm{3R}}$ of at least
\begin{equation*}
\nu_{\mathrm{3R}} \geq \frac{1}{n} \left[ |\region[S|\epsilon/2]| + \frac{\beta}{\alpha+\beta} \sum_{t \in \region[M|\epsilon/2]} (1-q_t) \right],
\end{equation*}
for a $0 < q_t \leq 1$, $t \in \region[M|\epsilon/2]$, as specified in \cref{sec:3r_mechanism}.
\end{theorem}

Furthermore, we construct a baseline, referred to as \ac{mq} mechanism, which mimics pre-existing approaches to designing mechanisms for correlated data. Note that those approaches are built around the idea of \di perturbations.
We give an upper bound on the utility of any \di local redaction mechanism and show that the \ac{mq} mechanism (\cref{algo:deterministic_redaction}) achieves this upper bound asymptotically (in $n$). The upper bound is given in \cref{thm:deterministic_redaction}. The proof is technical, and thus, deferred to \cref{apx:proof_deterministic_redaction}.
Herein, we observe that, depending on the privacy budget $\epsilon$, the optimal strategy is either to redact symmetrically around $X_\p$ or redact one side of the Markov chain completely and release only records from the other side.

\begin{theorem}
\label{thm:deterministic_redaction}
Let $\Delta^\star(\epsilon)$ denote the smallest integer such that $\infl{X_\p \rightsquigarrow X_{\p+\Delta^\star(\epsilon)}} \leq \epsilon$.
For different values of $\epsilon \geq 0$, the utility $\nu_\mathrm{DIM}$ of any $\epsilon$-private \di local redaction mechanisms is bounded from above by

\begin{align*}
    \nu_\mathrm{DIM} \leq \begin{cases}
        0 & \! \epsilon < \infl{X_\p \rightsquigarrow X_n},\\
        1 - \frac{R_2}{n} & \epsilon \geq \!\! \substack{\infl{X_\p \rightsquigarrow X_1}\\\quad+\infl{X_\p \rightsquigarrow X_n}},\\
        1-\frac{R_1}{n} & \! \text{otherwise},
    \end{cases}
\end{align*}
with $R_1 = \Delta^\star(\epsilon)+\p-1$, $R_2 = \min\{ R_1, 2 \Delta^\star(\epsilon/2)-1 \}$.

\end{theorem}

\begin{corollary}
    \label{cor:3r_improve_over_DI}
    \ac{3r} mechanisms always outperform the \ac{mq} mechanism. As a result, \ac{3r} mechanisms outperform any \di mechanism asymptotically (in $n$).
    \begin{proof}
        We can always find a \ac{3r} mechanism with utility $\nu_\mathrm{3R}$ at least the utility of the \ac{mq} mechanism, which is asymptotically optimal. The details are deferred to \cref{proof:3r_improve_over_DI}.
    \end{proof}
\end{corollary}

The utility gain of \ac{3r} mechanisms is illustrated in \cref{fig:deterministic_vs_randomized} for particular instantiations of \ac{3r} mechanisms explained in~\cref{sec:discussion}. This gain is ascribed to the exploitation of a more granular \dd leakage measure, which we call the pointwise-influence. Our results show that \di mechanisms are sub-optimal in a correlated data setting and demonstrate how \dd leakage information can help in designing good mechanisms.

\begin{figure}[t]
    \centering
    \resizebox{0.9\linewidth}{!}{\input{privacy-utility-fixed_p1_n10}}
    \caption{Comparison between the utility upper bound for \di mechanisms (DIM-UB), cf. \cref{thm:deterministic_redaction}, and the utility of two \ac{3r} mechanisms: a mechanism based on convex relaxation and a numerically optimized mechanism. The parameters are $\p=1$, $n=10$, $\alpha=0.01$, $\beta=0.8$.}
    \label{fig:deterministic_vs_randomized}
\end{figure}

\begin{example}[Motivating Example]
\label{eq:motivation_exampmle}
We illustrate how using the \dd pointwise-influence can increase the utility of a redaction mechanism for the same privacy budget.
Consider the simple example with $n=2$ and a correlation given by $\alpha=0.25$, $\beta=0.5$. The private record is $X_1$, and the privacy budget is $\epsilon=0.5$. From the likelihood ratios given in \cref{tab:lr} one can observe that $\infl{X_\p \rightsquigarrow \X_2} = \log(2)$, $\pwInfl{X_\p \rightsquigarrow X_2=0} = \log(3/2)$ and $\pwInfl{X_\p \rightsquigarrow X_2=1} = \log(2)$.
For any \di mechanism using the max-influence as a leakage measure, the records $X_1$ and $X_2$ must always be redacted since $\infl{X_\p \rightsquigarrow \X_2} = \log(2)>\epsilon = 0.5$. Thus, achieving a utility of $\nu_{\mathrm{DIM}}=0$. 

\begin{table}[h]
    \caption{Likelihood ratios $\Pr{X_2=x_2|X_1=x}/\Pr{X_2=x_2|X_1=\bar{x}}$ for different $x,x_2$ with $\alpha=0.25$, $\beta=0.5$.}
    \label{tab:lr}
    \centering
    \begin{tabular}{r|cc}
        & $x_2=0$ & $x_2=1$ \\
        \hline
        $x=0$ & 3/2 & 1/2 \\
        $x=1$ & 2/3 & 2
    \end{tabular}
\end{table}

Similarly, when considering \dd mechanisms using the pointwise-influence as a leakage measure, one must always redact $X_2$, when $X_2 = 1$ since $\pwInfl{X_\p \rightsquigarrow X_2=1} = \log(2)>\epsilon$. However, the main difference is that \dd mechanisms do not always have to redact $X_2$ when $X_2 = 0$. This is because $\pwInfl{X_\p \rightsquigarrow X_2=0} = \log(3/2)<\epsilon$. Nevertheless, to guarantee privacy, the mechanism cannot always release $X_2$, when $X_2 = 0$ since this deterministic output will always reveal the value of $X_2$, i.e., $0$ when it is released and $1$ when it is redacted. To avoid this artifact, $X_2$ should be redacted with a positive probability $q_2\define \Pr{Y_2=\perp|X_2=0}$. 
Guaranteeing $\epsilon$-privacy requires choosing $q_2$ such that the following holds
\begin{align*}
    \frac{ \Pr{X_2=\perp|X_1=1} }{ \Pr{X_2=\perp|X_1=0} } & = \frac{q_2 \beta + (1-\beta)}{\alpha + q_2 (1-\alpha)} \leq \exp(\epsilon).
\end{align*}
A privacy-preserving choice is $q_2=1/8$, which yields a utility
\(
    \nu_{\mathrm{DDM}}=\frac{1}{2} (1-q_2) \Pr{X_2=0}
    = 7/24
    \approx 0.292.
\)
That is, $\nu_{\mathrm{DDM}} > \nu_{\mathrm{DIM}}$.
\end{example}

\section{\acs{3r} Mechanisms}
\label{sec:3r_mechanism}

We introduce the family of \ac{3r} mechanisms and show how pointwise-influence helps improve the utility of redaction mechanisms.
A \ac{3r} mechanism separates the records into three sets $\recordS,\recordM,\recordL \subseteq \range{n}$, which are derived from the regions $\region[S|\epsilon^\prime]$, $\region[M|\epsilon^\prime]$, and $\region[L|\epsilon^\prime]$ defined in \cref{sec:setting}.
For any (but fixed) $\epsilon_\ell,\epsilon_r > 0$ such that $\epsilon_\ell+\epsilon_r \leq \epsilon$, define
\begin{align*}
\recordS &\define {\region[S|\epsilon_\ell]}^{(\ell)} \cup \region[S|\epsilon_r]^{(r)},\\
\recordM &\define \region[M|\epsilon_\ell]^{(\ell)} \cup \region[M|\epsilon_r]^{(r)},\\
\recordL &\define \region[L|\epsilon_\ell]^{(\ell)} \cup \region[L|\epsilon_r]^{(r)}.
\end{align*}
A record $X_t$ is said to be in region $\recordS$ ($\recordM$, $\recordL$, respectively) if $t \in \recordS$. Records in region $\recordL$ cause large leakage and, thus, need to be always redacted, i.e., $Y_t=\perp$ for $t \in \recordL$.
Records in region $\recordM$ cause a medium leakage and can be released%
\footnote{For $\alpha \leq \beta$, the pointwise-influence of $X_t=1$ is always larger than the pointwise-influence of $X_t=0$. For $\alpha > \beta$, the opposite is true.}
if $X_t=0$, but need to be redacted if $X_t=1$.
Records in region $\recordS$ cause small leakage and, thus, are allowed to be always released, i.e., $Y_t=X_t$ for $t \in \recordS$.
A \ac{3r} mechanism chooses to always redact records in $\recordL$, always release the records in $\recordS$, and ensures privacy by balancing the redactions in region $\recordM$, i.e.,
\begin{equation}
    \label{eq:mechanism_restriction}
    \Pr{Y_t=\perp | X_t = x_t} =
    \begin{cases}
        1 & t \in \recordL, \\
        0 & t \in \recordS, \\
        q_t & x_t = 0 \text{ and } t \in \recordM,\\
        1 & x_t = 1 \text{ and } t \in \recordM,
    \end{cases}
\end{equation}
for some $0 < q_t \leq 1$.
Thus, a \ac{3r} mechanism is determined by the choice of $\epsilon_\ell,\epsilon_r$ and by the choice of $q_t$.
The values $q_t$, $t \in \recordM$ are chosen such that
\begin{equation}
    \label{eq:3r_privacy_statement}
    \leak{X_\p \to \Y^{(\ell)}} \leq \epsilon_\ell,
    \quad
    \leak{X_\p \to \Y^{(r)}} \leq \epsilon_r.
\end{equation}
We remark that $\recordM$ can be empty, e.g., when $\alpha=\beta$. In such cases, a \ac{3r} mechanism cannot improve over \di mechanisms.
The size $|\recordM|$ depends on the parameters $\alpha,\beta$ and the privacy budget $\epsilon$.

\subsection{Privacy and utility}

The \ac{ldp} leakage caused by the left release $\Y^{(\ell)}$ and right release $\Y^{(r)}$ compose additively (cf. \cref{rmk:markovian_infl}). Therefore, it holds
\begin{equation*}
    \leak{X_\p \to \Y}
    \leq \leak{X_\p \to \Y^{(\ell)}} + \leak{X_\p \to \Y^{(r)}}
    \leq \epsilon,
\end{equation*}
where the last inequality holds by the condition in \cref{eq:3r_privacy_statement}. Thus, $\epsilon$-privacy is guaranteed.
The expected number of redacted records is
\begin{align*}
&\Exp{ \sum_{t=1}^{n} \indicator{X_t \neq Y_t} }
\\
=&|\recordL|+\sum_{t \in \recordM} \Pr{X_t=1} \cdot 1 +\sum_{t \in \recordM} \Pr{X_t=0} \cdot q_t
\\
=& |\recordL|+|\recordM| \frac{\alpha}{\alpha+\beta} + \frac{\beta}{\alpha+\beta} \sum_{t \in \recordM} q_t
\\
=& |\recordL|+|\recordM| - \frac{\beta}{\alpha+\beta} \sum_{t \in \recordM} (1-q_t),
\end{align*}
such that the utility is
\begin{align}
\nonumber
\nu_{\mathrm{3R}} &= 1-\frac{1}{n} \Exp{ \sum_{t=1}^{n} \indicator{X_t \neq Y_t}}
\\
\label{eq:3r_general_utility}
&= \frac{1}{n} \left[ |\recordS| + \frac{\beta}{\alpha+\beta} \sum_{t \in \recordM} (1-q_t) \right].
\end{align}
We present this result in \cref{thm:randomized_redaction} for the case $\epsilon_\ell=\epsilon_r=\epsilon/2$, serving as a stepping stone. The further increase of utility through an optimization of these parameters is left for future work. Hence, we obtain $\recordS=\region[S|\epsilon/2]$ and $\recordM=\region[M|\epsilon/2]$ and can express the achievable utility as
\begin{equation*}
\nu_{\mathrm{3R}} = \frac{1}{n} \left[ |\region[S|\epsilon/2]| + \frac{\beta}{\alpha+\beta} \sum_{t \in \region[M|\epsilon/2]} (1-q_t) \right].
\end{equation*}

\subsection{Mechanism design}
\label{sec:3r_example}
\balance

We finally consider a particular \ac{3r} mechanism design approach, which yields a simple closed-form solution. 
For the ease of presentation, we focus on the case $\p=1$ with $\epsilon_r=\epsilon$. For $\p>1$, the same solution is applicable to find respective redaction probabilities for the records left and right of $X_\p$.

Expanding the privacy condition in \cref{def:ldp}, it is possible to bound
\begin{equation}
    \label{eq:loose_bound}
    \leak{\X_p \to \Y}
    \leq
    \max_{t \in \recordM} \left( \delta_t - \sum_{i \in \recordM_t} \log(q_i) \right),
\end{equation}
where $\recordM_t \define \recordM \cap \{ i \colon |i-\p| \leq |t-\p| \}$ and
\begin{equation*}
    \delta_t
    \define
    \begin{cases}
        0 & t+1 > n,\\
        \pwInfl{X_\p \rightsquigarrow X_{t+1}=0} & t+1 \in \recordM,\\
        \pwInfl{X_\p \rightsquigarrow X_{t+1}=1} & t+1 \in \recordS.
    \end{cases}
\end{equation*}
We defer the detailed technical derivation of this bound to \cref{apx:privacy_conditions}.
Maximizing the utility under the constraint in \cref{eq:loose_bound} can be formulated as a convex optimization problem, which can be solved efficiently, e.g., by interior point methods~\cite{boyd2004convex}. For various parameter choices, we observed that the optimal solution is of the form $q_t=q$ for all $t \in \recordM$. Hence, for clarity of exposition, we give the solution under this assumption, which is
\begin{equation*}
    q_t = \max_{i \in \recordM} \exp\left( - (\epsilon-\delta_i)/|\recordM_i| \right).
\end{equation*}
The full optimization problem is stated in \cref{apx:privacy_conditions}.
By definition of $\recordM$ and $\recordS$, it holds $\delta_t \leq \epsilon$. Thus, we always obtain values $0 \leq q_i \leq 1$.

\section{Discussion and Conclusion}
\label{sec:discussion}

We numerically evaluate the utility achievable by \ac{3r} mechanisms and compare it to the utility upper-bound of \di mechanisms according to \cref{thm:deterministic_redaction}. The utility is given as a function of the privacy budget $\epsilon$ in \cref{fig:deterministic_vs_randomized}, where the parameters are $\p=1$, $n=10$, $\alpha=0.01$, $\beta=0.8$. We evaluate two different \ac{3r} mechanism designs: first, the mechanism from \cref{sec:3r_example}, which is based on a relaxed leakage bound; and second, a mechanism that uses numerically optimized redaction probabilities in $\recordM$. For the latter, we define $q_t=q$ for all $t \in \recordM$ and perform a grid search for feasible values for $q$. We also depict the resulting redaction probabilities $\Pr{Y_t=\perp|X_t=0}$ in \cref{fig:random_redaction_profile} for a privacy budget $\epsilon=1$.

\begin{figure}[t]
    \centering
    \resizebox{0.9\linewidth}{!}{\begin{tikzpicture}

\begin{axis}[%
width=\linewidth,
height=0.6\linewidth,
scale only axis,
xmin=1,
xmax=5,
xtick=data,
xlabel={Record index $t$},
ymin=-0.1,
ymax=1.1,
ylabel={$\Pr{Y_t=\perp|X_t=0}$},
axis background/.style={fill=white},
legend style={legend cell align=left, align=left, at={(0.01,0.01)}, anchor=south west},
grid=both,
x grid style={line width=.1pt, draw=black!10, dashed},
major grid style={line width=.2pt,draw=black!20},
minor grid style={draw=none},
minor tick num=1,
]

\addplot [color=mq,mark=asterisk,thick]
  table[row sep=crcr]{%
1	1\\
2	1\\
3	1\\
4	0\\
5	0\\
6	0\\
7	0\\
8	0\\
9	0\\
10	0\\
};
\addlegendentry{MQM}

\addplot [color=3rConvex,mark=o,thick]
  table[row sep=crcr]{%
1	1\\
2	0.757414764826369\\
3	0.757414731740435\\
4	0\\
5	0\\
6	0\\
7	0\\
8	0\\
9	0\\
10	0\\
};
\addlegendentry{3R (relaxation)}

\addplot [color=3rNumerical,mark=*,thick]
  table[row sep=crcr]{%
1	1\\
2	0.547547547547548\\
3	0.547547547547548\\
4	0\\
5	0\\
6	0\\
7	0\\
8	0\\
9	0\\
10	0\\
};
\addlegendentry{3R (numerical)}

\end{axis}
\end{tikzpicture}%
}
    \caption{Redaction probabilities employed by 
    different mechanisms for a privacy budget $\epsilon=1$ and parameters $\p=1$, $n=10$, $\alpha=0.01$, $\beta=0.8$.}
    \label{fig:random_redaction_profile}
\end{figure}

The numerical results demonstrate that \ac{3r} mechanisms outperform \di local redaction mechanisms when the redaction probabilities $q_t$ are designed appropriately. While the relaxation-based mechanism improves the utility for large privacy budgets in particular, its advantage decreases for small privacy budgets. However, the utility for numerically optimized redaction probabilities demonstrates the general potential of \ac{3r} mechanisms, even under restriction to constant redaction probabilities.
A \ac{3r} mechanism gains in utility by reducing the redaction probabilities in the medium leakage region $\region[M|\epsilon^\prime]$.
Therefore, the gain depends on this region's size, which grows with $\beta/\alpha$.
Discontinuities in \cref{fig:deterministic_vs_randomized} are due to discontinuities in the size of this region.
Overall, our results show that \di mechanisms are sub-optimal in correlated data settings, and demonstrate how the pointwise-influence helps in designing good mechanisms.

As future work, we aim to generalize our ideas to more complex data distributions and correlation models. Furthermore, we will investigate the optimal privacy-utility tradeoff among all \ac{3r} mechanisms, which includes the optimal privacy budget split and redaction probability design.

\newpage
\balance
\bibliographystyle{ieeetr}

\newpage
\appendix
\subsection{Detailed derivation of relaxed 3R privacy conditions}
\label{apx:privacy_conditions}

\begin{figure*}[t]
\begin{align}
\label{eq:privacy_condition_r}
\log\frac{\Pr{X_{t_+}=x_{t_+} | X_\p=x}}{\Pr{X_{t_+}=x_{t_+} | X_\p=\bar{x}}}
+ \log \frac{
    \sum_{\mathcal{S} \in \powerset(\recordM_t^{(r)})}
        \left( \prod_{i \in \mathcal{S}} q_i \right)
        \Pr{ \X_{\mathcal{S}}=\zero, \X_{\recordM_t^{(r)} \setminus \mathcal{S}}=\one | X_\p=x, X_{t_+}=x_{t_+} }
}{
    \sum_{\mathcal{S} \in \powerset(\recordM_t^{(r)})}
        \left( \prod_{i \in \mathcal{S}} q_i \right)
        \Pr{ \X_{\mathcal{S}}=\zero, \X_{\recordM_t^{(r)} \setminus \mathcal{S}}=\one | X_\p=\bar{x}, X_{t_+}=x_{t_+} }
}
& \leq \epsilon_r
\\
\label{eq:privacy_condition_full_redaction}
 \log \frac{
    \sum_{\mathcal{S} \in \powerset(\recordM^{(r)})}
        \left( \prod_{i \in \mathcal{S}} q_i \right)
        \Pr{ \X_{\mathcal{S}}=\zero, \X_{\recordM^{(r)} \setminus \mathcal{S}}=\one | X_\p=x }
}{
    \sum_{\mathcal{S} \in \powerset(\recordM^{(r)})}
        \left( \prod_{i \in \mathcal{S}} q_i \right)
        \Pr{ \X_{\mathcal{S}}=\zero, \X_{\recordM^{(r)} \setminus \mathcal{S}}=\one | X_\p=\bar{x} }
}
& \leq \epsilon_r
\end{align}
\end{figure*}

Since the forward and backward transition probabilities are the same in our setting, both expressions $\leak{X_\p \to \Y^{(\ell)}}$ and $\leak{X_\p \to \Y^{(r)}}$ can be written in the form $\leak{\tilde{X}_1 \to \tilde{\Y}_{[1,\tilde{n}]}}$ under re-indexing. Hence, it is sufficient to consider the case $\p=1$ and show that
\begin{equation}
    \label{eq:privacy_definition_pointwise}
    \frac{\Pr{ \Y = \y | X_\p=x }}{\Pr{ \Y = \y | X_\p=\bar{x} }} \leq \exp(\epsilon)
\end{equation}
for every mechanism output $\y$ that occurs with non-zero probability and for every $x \in \mathcal{X}$.

Suppose that there is a non-redacted record in the output.
According to \cref{eq:mechanism_restriction}, every such mechanism output in this case can be written as
\begin{equation*}
    \y^{(t)}=(\perp,\dots,\perp,x_{t+1},\y_{[t+2,n]})
\end{equation*}
where $t \in \recordM \cup \{\max\recordL\}$ and $x_{t+1} \in \supp(Y_{t+1}) \cap \mathcal{X}$.
When $t = \max\recordL$, then $t+1 \in \recordM$ and
\begin{align*}
    &\frac{\Pr{ \Y = \y^{(t)} | X_\p=x }}{\Pr{ \Y = \y^{(t)} | X_\p=\bar{x} }}
    = \frac{\Pr{X_{t+1}=x_{t+1} | X_1=x}}{\Pr{X_{t+1}=x_{t+1} | X_1=\bar{x}}}
    \\
    \leq& \exp( \pwInfl{X_\p \rightsquigarrow X_{t+1}=x_{t+1}} ) \leq \exp(\epsilon_r)
\end{align*}
holds by definition of $\recordM$. Hence, \cref{eq:privacy_definition_pointwise} is always satisfied for $t = \max\recordL$.
For $t \in \recordM$ we can write
\begin{align*}
    &\Pr{ \Y = \y^{(t)} | X_\p=x }
    \\
    =& \sum_{\x_{\recordM_t}}
    \Pr{ \X_{\recordM_t} =\x_{\recordM_t}, X_{t+1}=x_{t+1} | X_1=x } \prod_{\substack{i \in {\recordM_t}\\x_i=0}} p_i
    \\
    & \quad \cdot \Pr{ \Y_{[t+2,n]}=\y_{[t+2,n]} | X_{t+1}=x_{t+1} }
    \\
    =& \,\Pr{ \Y_{[t+2,n]}=\y_{[t+2,n]} | X_{t+1}=x_{t+1} }
    \\
    & \cdot \Pr{X_{t+1}=x_{t+1} | X_1=x}
    \\
    & \cdot \sum_{\x_{\recordM_t}}
    \Pr{ \X_{\recordM_t} =\x_{\recordM_t} | X_{t+1}=x_{t+1}, X_1=x } \prod_{\substack{i \in {\recordM_t}\\x_i=0}} p_i.
\end{align*}
Using this expression, we can upper-bound \cref{eq:privacy_definition_pointwise} as \cref{eq:privacy_condition_r}, where we define $t_+=t+\operatorname{sign}(t-\p)$ and $\recordM_t \define \recordM \cap \{ i \colon |i-\p| \leq |t-\p| \}$  (here, $\recordM^{(r)}=\recordM$).

Finally, consider the output $\y^{(\perp)}=(\perp,\dots,\perp)$, which applies only in case $\region[S|\epsilon_r]=\emptyset$. Then \cref{eq:privacy_condition_full_redaction} holds since
\begin{align*}
    &\Pr{ \Y = \y^{(\perp)} | X_\p=x }
    \\
    =& \sum_{\x_{\recordM}}
        \Pr{ \X_{\recordM}=\x_{\recordM} | X_\p=x }
        \prod_{\substack{i \in {\recordM}\\x_i=0}} p_i,
\end{align*}

Overall, privacy is satisfied if the values $q_t$ are such that \cref{eq:privacy_condition_r} and \cref{eq:privacy_condition_full_redaction} hold
for every $x \in \mathcal{X}$, $x_{t+1} \in \supp(Y_{t+1}) \cap \mathcal{X}$, and for every $t \in \recordM^{(r)}$ (to be precise, \cref{eq:privacy_condition_r} holds for $t \in \recordM^{(r)} \setminus \{n\}$  and \cref{eq:privacy_condition_full_redaction} holds for $t \in \recordM^{(r)} \cap \{n\}$).
We can bound

\begin{align*}
    &\text{LHS of } (\ref{eq:privacy_condition_r})
    \\
    \leq &
    \pwInfl{X_\p \rightsquigarrow X_{t_+}=x_{t_+}}
    + \log\frac{\min_\mathcal{S} \prod_{i \in \mathcal{S}} q_i}{\max_\mathcal{S} \prod_{i \in \mathcal{S}} q_i}
    \\
    \leq & 
    \max_{x_{t_+} \in \supp(Y_{t_+}) \cap \mathcal{X}} \pwInfl{X_\p \rightsquigarrow X_{t_+}=x_{t_+}}
    - \sum_{i \in \recordM_t^{(r)}} \log(q_i),
\end{align*}
and accordingly,
\begin{align*}
    &\text{LHS of } (\ref{eq:privacy_condition_full_redaction}
    \leq &
    \log\frac{\min_\mathcal{S} \prod_{i \in \mathcal{S}} q_i}{\max_\mathcal{S} \prod_{i \in \mathcal{S}} q_i}
    \leq & 
    - \sum_{i \in \recordM^{(r)}} \log(q_i).
\end{align*}
Combining these bounds gives:
\begin{equation*}
    \leak{\X_p \to \Y}
    \leq
    \max_{t \in \recordM^{(r)}} \left( \delta_t - \sum_{i \in \recordM^{(r)}_t} \log(q_i) \right),
\end{equation*}
where $\recordM_t \define \recordM \cap \{ i \colon |i-\p| \leq |t-\p| \}$ and
\begin{equation*}
    \delta_t
    \define
    \begin{cases}
        0 & t_+ \notin [1,n],\\
        \pwInfl{X_\p \rightsquigarrow X_{t_+}=0} & t_+ \in \recordM,\\
        \pwInfl{X_\p \rightsquigarrow X_{t_+}=1} & t_+ \in \recordS.
    \end{cases}
\end{equation*}

Thus, a \ac{3r} mechanism is $\epsilon$-private if $\delta_t - \sum_{i \in \recordM_t^{(r)}} \log(q_i) \leq \epsilon_r$ for all $t \in \recordM^{(r)}$. By the concavity of the logarithm, these constraints are convex in $q_t$. The utility of a \ac{3r} mechanism is strictly increasing in $\sum_{t \in \recordM^{(r)}} q_t$, cf.~\cref{eq:3r_general_utility}, which is a linear function in $q_t$. Hence, we can efficiently optimize the values $q_t$ for the above bound by solving the convex optimization problem:
\begin{align*}
    \min_{q_t,\, t \in \recordM^{(r)}}& \sum_{t \in \recordM^{(r)}} q_t\\
    \text{subject to }& - \sum_{i \in \recordM^{(r)}_t} \log(q_i) \leq \epsilon_r-\delta_t \, \forall_{t \in \recordM^{(r)}}\\
    & 0 \leq q_t \leq 1 \, \forall_{t \in \recordM^{(r)}}
\end{align*}
By definition of $\recordM^{(r)}$ and $\recordS^{(r)}$, it holds $\delta_t \leq \epsilon_r$. Thus, there always exists a feasible solution with $0 \leq q_i \leq 1$. As an example, \cref{fig:random_redaction_profile} compares the resulting redaction probabilities of this design approach against the ones for numerically optimized $q$ such that $q_t=q$ and for the \ac{mq} mechanism.

When $\p>1$, the same derivations apply to $q^{(\ell)}=q_t$ for $t \in \recordM^{(\ell)}$, where $\recordM^{(r)}$ and $\epsilon_r$ are substituted by $\recordM^{(\ell)}$ and $\epsilon_\ell$, respectively.

\subsection{3R mechanisms improve over \di mechanisms asymptotically}
\label{proof:3r_improve_over_DI}

We argue that a \ac{3r} mechanism can have higher utility than the \ac{mq} mechanism (\cref{algo:deterministic_redaction}). Since the \ac{mq} mechanism is an asymptotically (in $n$) optimal \di mechanism, we conclude that, asymptotically, there exists a \ac{3r} mechanism with higher utility than every \di mechanism.

Depending on the privacy budget $\epsilon$ and the parameters $n,\p$, the \ac{mq} mechanism chooses either $\epsilon_\ell=\epsilon_r=\epsilon/2$ or $\epsilon_\ell=0,\epsilon_r=\epsilon$.
In both cases, $\epsilon_\ell+\epsilon_r=\epsilon$ holds and $\Pr{Y_t=\perp}<1$ only if $X_t \in \recordS$.

Hence, the utility of the \ac{mq} mechanism can be given as $\nu_\mathrm{MQ} \leq |\recordS|/n$, for $\recordS$ as defined in \cref{sec:3r_mechanism}.
For the same parameters $\epsilon_\ell,\epsilon_r$, there exists a \ac{3r} mechanism with utility given by \cref{eq:3r_general_utility} as
\begin{equation*}
    \nu_\mathrm{3R} = 
    \frac{1}{n} \left[ |\recordS| + \frac{\beta}{\alpha+\beta} \sum_{t \in \recordM} (1-q_t) \right] \geq \nu_\mathrm{MQ}.
\end{equation*}

\subsection{The Markov Quilt mechanism}
\label{sec:mq_mechanism}

In~\cite{songPufferfishPrivacyMechanisms2017a}, a privacy mechanism was proposed that operates purely based on the max-influence. Although the original mechanism employs perturbation, and thus, is not directly applicable to our setting, we translate the main idea to the redaction setting and consider it as a baseline. Following the nomenclature of~\cite{songPufferfishPrivacyMechanisms2017a}, we call it the \ac{mq} mechanism. In addition, we study the family of \di local redaction mechanisms, which the \ac{mq} mechanism is part of. 

For a private $X_\p$, the mechanism in~\cite{songPufferfishPrivacyMechanisms2017a} searches for a so-called Markov quilt $\mathcal{Q} \subset \range{n}$, which is a set of records surrounding $X_\p$, such that its max-influence $\infl{X_\p \rightsquigarrow \X_{\mathcal{Q}}}$ is sufficiently small. Perturbations are then tuned according to a function of $\infl{X_\p \rightsquigarrow \X_{\mathcal{Q}}}$.
\begin{definition}[Markov quilt]
\label{def:markov_quilt}
A set $\mathcal{Q} \subset \range{n}$ with records $\X_\mathcal{Q}$
is a Markov quilt for a record $X_\p$ if
\begin{itemize}
    \item there exists a partition $\{ \mathcal{Q}, \mathcal{N}, \mathcal{R} \}$ of $\range{n}$ with $X_\p \in \X_\mathcal{N}$ (\emph{nearby} records),
    \item for all $\x \in \mathcal{X}^n$ the records in $\X_\mathcal{R}$ (\emph{remote} records) are conditionally independent from the records in $\X_\mathcal{N}$ given $\X_\mathcal{Q}$, i.e.,
     \begin{align*}
     & \Pr{ \X_\mathcal{R} = \x_\mathcal{R} | \X_\mathcal{Q}=\x_\mathcal{Q}, \X_\mathcal{N}=\x_\mathcal{N}} \\
     & = \Pr{ \X_\mathcal{R} = \x_\mathcal{R} | \X_\mathcal{Q}=\x_\mathcal{Q}}.
     \end{align*}
\end{itemize}
\end{definition}

For our setting, the max-influence determines the set of redacted records instead. Since the max-influence is decreasing in $|\p-t|$, an intuitive solution is to build a redaction window around $X_\p$, which spans nearby records $\X_\mathcal{N}$. Records outside of the redaction window form a Markov quilt that dictates the privacy leakage.

\begin{algorithm}[ht]
    \SetAlgoLined
    \SetKwInOut{Input}{Input}
    \SetKwInOut{Output}{Output}
    \SetKwInOut{Require}{Require}

    \Input{$\epsilon > 0$, $n$, $\x \in \mathcal{X}^n$, $0 \leq \p \leq n/2$}
    $\Delta_\epsilon \gets \Delta^\star(\epsilon)$ ;
    $\Delta_{\epsilon/2} \gets \Delta^\star(\epsilon/2)$ \;
    \uIf{$\p=1 \,\lor\, \epsilon < \highInfl{n+1-\p}+\highInfl{\p-1} \,\lor\, t_\p(\epsilon) < 0$}{
        $\Delta_\ell \gets \p-1$ ;
        $\Delta_r \gets \min\{ \Delta_\epsilon, n-\p \}$ \;
    }
    \Else{
        $\Delta_\ell \gets \Delta_{\epsilon/2}$ ;
        $\Delta_r \gets \Delta_{\epsilon/2}$ \;
    }
    $\mathcal{N} \gets [\p-\Delta_\ell,\p+\Delta_r]$ \;
    $\y \gets \x$;
    $\y_\mathcal{N} \gets \perp$\;
    \Output{$\y$}
    \caption{Markov quilt redaction mechanism.}
    \label{algo:deterministic_redaction}
\end{algorithm}

\paragraph*{Privacy and utility of the \ac{mq} mechanism}
We can show that the \ac{mq} mechanism achieves a utility $\nu_\mathrm{MQ}$ close to the upper bound in \cref{thm:deterministic_redaction}:
\begin{align*}
    \nu_\mathrm{MQ} \geq \begin{cases}
        0 & \! \epsilon < \infl{X_\p \rightsquigarrow X_n},\\
        1 - \frac{\min\{R_1,R_2\}}{n} - \frac{2}{n}& \! \substack{\epsilon \geq \infl{X_\p \rightsquigarrow X_1}\\\quad+\infl{X_\p \rightsquigarrow X_n}},\\
        1-\frac{R_1}{n} - \frac{1}{n} & \! \text{otherwise},
    \end{cases}
\end{align*}
with $R_1 = \Delta^\star(\epsilon)+\p-1$ and $R_2 = 2 \Delta^\star(\epsilon/2)-1$. That is, it is asymptotically optimal (in $n$). The derivation is straight-forward by expressing the utility as $\nu_\mathrm{MQ} = 1-\frac{\Delta_\ell+\Delta_r+1}{n}$ and using $\Delta_\ell$ and $\Delta_r$ as given in \cref{algo:deterministic_redaction} for the three different cases.
The privacy follows directly from the definition of $\Delta^\star(\epsilon)$ and $\Delta^\star(\epsilon/2)$, respectively, and from the composition rule in \cref{rmk:markovian_infl} given in \cref{apx:proof_deterministic_redaction}.

\subsection{Proof of \cref{thm:deterministic_redaction}}
\label{apx:proof_deterministic_redaction}

The leakage of a \di local redaction mechanism is determined by a Markov quilt that surrounds the redaction window as stated by \cref{cor:leakage_bound}. Thus, the proof idea is to consider those Markov quilts that lead to a leakage of at most $\epsilon$ as candidates. Among these candidates, we find the Markov quilt that maximizes the utility. To simplify the notation, we use the notion of a \emph{redaction radius}. We define the set of released records as
\begin{equation}
    \label{eq:set_of_releases}
    \region[rel] = \range{n} \cap \{ t \colon \Pr{Y_t=\perp}<1 \}.
\end{equation}

\begin{definition}[Redaction radius]
\label{def:redaction_radius}
For a \di mechanism, with $\region[rel]$ as defined in \cref{eq:set_of_releases}, the redaction radius for a private record $X_\p$ is defined as
\(
    \rr = (\Delta_\ell,\Delta_r)
\)
where
\begin{itemize}
    \item $\Delta_\ell$ is the largest integer such that $j \notin \region[rel]$ for all \mbox{$j \geq \p-\Delta_\ell$},
    \item $\Delta_r$ is the largest integer such that $j \notin \region[rel]$ for all \mbox{$j \leq \p+\Delta_r$}.
\end{itemize}
\end{definition}

Furthermore, we can give an exact composition rule for the leakage of \di mechanisms.
\begin{remark}[Composition for \di mechanisms]
    \label{rmk:markovian_infl}
    Let $\mathcal{S}^{(\ell)} = \mathcal{S} \cap [1,t_\ell]$, $\mathcal{S}^{(r)} = \mathcal{S} \cap [t_r,n]$ be a partition of $\mathcal{S} \subseteq \range{n}$ with $t_\ell,t_r \in \mathcal{S}$ and $t_\ell \leq \p \leq t_r$. By the Markovity of the correlation, it then holds for $t_r-t_\ell$ even,
    \begin{equation*}
        \infl{X_\p \rightsquigarrow \X_\mathcal{S}}
        =
        \infl{X_\p \rightsquigarrow \X_{\mathcal{S}^{(\ell)}}}
        +
        \infl{X_\p \rightsquigarrow \X_{\mathcal{S}^{(r)}}}.
    \end{equation*}
    As a consequence from \cref{cor:leakage_bound}, a \di local redaction mechanism has
    \begin{equation*}
        \leak{X_\p \to \Y}
        =
        \leak{X_\p \to \Y_{\mathcal{S}^{(\ell)}}}
        +
        \leak{X_\p \to \Y_{\mathcal{S}^{(r)}}}.
    \end{equation*}
    for $t_r-t_\ell$ even. For general $t_\ell,t_r$ we have
    \begin{equation*}
        \infl{X_\p \rightsquigarrow \X_\mathcal{S}}
        \leq
        \infl{X_\p \rightsquigarrow \X_{\mathcal{S}^{(\ell)}}}
        +
        \infl{X_\p \rightsquigarrow \X_{\mathcal{S}^{(r)}}}.
    \end{equation*}
    and
    \begin{equation*}
        \leak{X_\p \to \Y}
        \leq
        \leak{X_\p \to \Y_{\mathcal{S}^{(\ell)}}}
        +
        \leak{X_\p \to \Y_{\mathcal{S}^{(r)}}}.
    \end{equation*}
\end{remark} 
To also relate the privacy leakage to the redaction radius, we translate \cref{cor:leakage_bound} into \cref{cor:deterministic_redaction_leakage}.
\begin{corollary}
    \label{cor:deterministic_redaction_leakage}
    Let $\Delta_\ell+\Delta_r$ even or $\Delta_\ell=\p-1$. A mechanism with deterministic redactions $\region[red]$ and $\rr=(\Delta_\ell,\Delta_r)$ is $\epsilon$-private only if
    \begin{equation*}
        \epsilon \geq \indicator{\Delta_\ell<\p-1} \highInfl{\Delta_\ell+1} + \indicator{\Delta_r<n-\p} \highInfl{\Delta_r+1}
    \end{equation*}
    \begin{proof}
        Define a suitable Markov quilt $\mathcal{Q} \subseteq \range{n} \setminus \R_{det}$ according to the redaction radius $\rr = (\Delta_\ell,\Delta_r)$ next. Let $\mathcal{Q}=\mathcal{Q}^{(\ell)} \cup \mathcal{Q}^{(r)}$ with
        \begin{align*}
            \mathcal{Q}^{(\ell)}
            &=
            \begin{cases}
            \emptyset & \Delta_\ell=\p-1,\\
            \{X_{\p-(\Delta_\ell+1)}\} & \text{otherwise},
            \end{cases}
            \\
            \mathcal{Q}^{(r)}
            &=
            \begin{cases}
            \emptyset & \Delta_r=n-\p\\
            \{X_{\p+(\Delta_r+1)}\}, & \text{otherwise}.
            \end{cases}
        \end{align*}

        By \cref{cor:leakage_bound} it holds $\epsilon \geq \leak{X_\p \to \Y} = \infl{X_\p \rightsquigarrow \X_{\mathcal{Q}}}$. Furthermore, we can apply \cref{rmk:markovian_infl} and write
        \begin{align*}
        \infl{X_\p \rightsquigarrow \X_{\mathcal{Q}}} &=& \infl{X_\p \rightsquigarrow \X_{\mathcal{Q}^{(\ell)}}} + \infl{X_\p \rightsquigarrow \X_{\mathcal{Q}^{(r)}}}\\
        &=& \indicator{\Delta_\ell<\p-1} \infl{X_\p \rightsquigarrow X_{\p-(\Delta_\ell+1)}} \\
        &&+ \indicator{\Delta_r<n-\p} \infl{X_\p \rightsquigarrow X_{\p+(\Delta_r+1)}}.
        \end{align*}
        If $\Delta_\ell=\p-1$, then $\mathcal{Q} = \mathcal{Q}^{(r)}$, and the leakage is $\infl{X_\p \rightsquigarrow \X_{\mathcal{Q}}}=\infl{X_\p \rightsquigarrow \X_{\mathcal{Q}^{(r)}}}$ trivially. 
        In both cases, we use the functional representation $\highInfl{\Delta}$ for the max-influence as defined in \cref{prop:max_influence_by_distance} to arrive at the desired statement.
    \end{proof}
\end{corollary}

Motivated by this, we split our analysis into two parts. We first consider releases of records left and right from $X_\p$ independently, and find the optimal Markov quilt in a one-sided privacy problem, i.e., when $\p=1$. In this case, we use the privacy budget $\epsilon=\epsilon_r$ for releasing records from the right of $X_\p$. After that, we generalize to general $\p$ by discussing the optimal split of the privacy budget $\epsilon$ onto releases from the right ($\epsilon_r$) and left ($\epsilon_\ell$) of $X_\p$, respectively. For the latter step, we apply \cref{cor:deterministic_redaction_leakage}, which only holds for even-sized redaction radii however. This introduces a small gap in the resulting lower bound that is negligible for large $n$. Recall the we assume $\p \leq n/2$ throughout the paper, such that the majority of records lives in the right chain.

\paragraph{One-sided redaction}
Suppose $\p=1$ and privacy budget $\epsilon_r$. Then the redaction radius is $\rr=(0,\Delta_r)=(\p-1,\Delta_r)$ and by \cref{cor:deterministic_redaction_leakage}, we require $\highInfl{\Delta_r+1} \leq \epsilon_r$.
Note that $\highInfl{\Delta}$ is a strictly decreasing function for $\Delta \in \mathbb{N}$. Hence, there exists a $\Delta^\star(\epsilon) \in \mathbb{N}$ such that $\highInfl{\Delta} \leq \epsilon$ implies $\Delta \geq \Delta^\star(\epsilon)$.
In our case, choosing $\Delta_r = \Delta^\star(\epsilon_r)-1$ minimizes the redaction radius (maximizes utility) for privacy budget $\epsilon_r$.

\paragraph{Two-sided redaction}
If $1 < \p < n$, the privacy budget epsilon can be split into $\epsilon_\ell$ and $\epsilon_r$ with $\epsilon_\ell+\epsilon_r=\epsilon$. In the following, we argue that the optimal redaction strategy depends on the privacy requirement $\epsilon$. Namely, for small $\epsilon$, utility is maximized by redacting the left Markov chain completely, and balancing the leakage through the redaction radius $\Delta_r$ in the right Markov chain. For sufficiently large $\epsilon$, utility is maximized by symmetrically redacting from both sides with $\Delta_\ell-1 \leq \Delta_r \leq \Delta_\ell+1$.
We distinguish these cases by a threshold $t_i(\epsilon)$.

Define $\epsilon_0 = \highInfl{n+1-i} < \highInfl{n-i}$ and $\epsilon_1 = \highInfl{i-1}$
By the assumption $i \leq n/2$ it holds that $i-1 \leq n-i < n+1-i$. Since $\highInfl{\Delta}$ is strictly decreasing in $\Delta$, we obtain that $\epsilon_0 < \epsilon_1$. Consider the following cases:

\underline{Case 0: $\epsilon < \epsilon_0$}
If $\epsilon < \epsilon_0 \leq \epsilon_1$, releasing any record will cause a leakage of $\leak{X_\p \to \Y} \geq \epsilon_0$, and thus, all records need to be redacted, i.e., $\nu_\mathrm{DIM} = 0$.

\underline{Case 1: $\epsilon_0 \leq \epsilon < \epsilon_1+\epsilon_0$}
If $\epsilon < \epsilon_1$, releasing any record from the left Markov chain will cause a leakage of $\leak{X_\p \to \Y} \geq \epsilon_1$. Hence, the left Markov chain must be redacted completely in this case: $\rr=(\p-1,\Delta^\star(\epsilon)-1)$.

If $\epsilon_1 \leq \epsilon < \epsilon_1+\epsilon_0$,
releasing records from the left or the right chain might be possible. If at the same time, record $X_\ell$ from the left chain and record $X_r$ from the right chain is released, then
\begin{align*}
    \leak{X_\p \to \Y} &\geq \leak{X_\p \to \Y_{\{\ell,r\}}} \geq \leak{X_\p \to \Y_{\{\ell,\tilde{r}\}}}
    \\
    &\overset{(a)}{=} \leak{X_\p \to \Y_{\ell}} + \leak{X_\p \to \Y_{\tilde{r}}}.
\end{align*} 
By $r \leq \tilde{r} \leq r+1$ we denote the smallest integer such that $\ell+\tilde{r}$ is even; thus, we can apply \cref{cor:deterministic_redaction_leakage} in $(a)$. However, $\leak{X_\p \to \Y_{\ell}} = \epsilon_1$ and $\leak{X_\p \to \Y_{\tilde{r}}} \geq \epsilon_0$, respectively. Hence, either the left Markov chain or the right Markov chain needs to be redacted completely: $\rr \in \{ (\p-1,\Delta_r^{(1)}), (\Delta_\ell^{(2)},n-\p) \}$.

As we assumed equal transition probabilities for the left and right Markov chain, the utility is maximized by the same value $\Delta_r^{(1)}=\Delta_\ell^{(2)}=\Delta^\star(\epsilon)-1$, respectively (see derivation for one-sided redaction).
However, since $\p-1 \leq n-\p$, the overall utility $\Delta_\ell+\Delta_r$ is maximized for $\rr = (\p-1,\Delta_r^{(1)}) = (\p-1,\Delta^\star(\epsilon)-1)$.
In summary, we bound the utility by
\begin{equation*}
    \nu_\mathrm{DIM} \leq 1 - \frac{1}{n} ( \p+\Delta^\star(\epsilon)-1 ).
\end{equation*}

\underline{Case 2: $\epsilon \geq \epsilon_1+\epsilon_0$}
The privacy requirement can be sufficiently large to allow for both; one-sided leakage $\rr=(\p-1,\Delta^\star(\epsilon)-1)$, or two-sided leakage $\rr=(\Delta_\ell,\Delta_r)$ with $\Delta_\ell < \p-1$ and $\Delta_r < n-\p$. 
In the case of one-sided leakage, the number of redacted records is at least
\begin{equation}
    \label{eq:rr_one-sided-leakage}
    \Delta_\ell+\Delta_r=\p+\Delta^\star(\epsilon)-2.
\end{equation}

Let $\tilde{\Delta}_r \geq \Delta_r$ denote the smallest integer such that $\Delta_\ell+\tilde{\Delta}_r$ is even. Then \cref{cor:deterministic_redaction_leakage} yields the privacy requirement $\highInfl{\Delta_\ell+1}+\highInfl{\tilde{\Delta}_r+1} \leq \epsilon$.
Let $\tilde{i}(\Delta)$ denote the affine extension of $\highInfl{\Delta}$.
By the convexity of the max-influence (\cref{prop:max_influence_by_distance}), we can apply Jensen's inequality
\begin{align*}
    & \frac{1}{2} \tilde{i}(\Delta_\ell+1) + \frac{1}{2} \tilde{i}(\tilde{\Delta}_r+1) \geq \tilde{i}(\frac{1}{2}(\Delta_\ell+1) + \frac{1}{2} (\tilde{\Delta}_r+1))
    \\
    \Leftrightarrow& \tilde{i}(\Delta_l+1)+\tilde{i}(\tilde{\Delta}_r+1) \geq 2 \tilde{i}(\frac{\Delta_\ell+\tilde{\Delta}_r}{2}+1).
\end{align*}
Hence, the privacy requirement is only satisfied if
\begin{equation*}
    \tilde{i}(\frac{\Delta_\ell+\tilde{\Delta}_r}{2}+1) \leq \epsilon/2.
\end{equation*}
Therefore, it holds that $1+\frac{\Delta_\ell+\tilde{\Delta}_r}{2} \geq \tilde{i}^{-1}(\epsilon/2)$, and since $\frac{\Delta_\ell+\tilde{\Delta}_r}{2}$ is an integer for $\Delta_\ell+\tilde{\Delta}_r$ even, we can write
\begin{align*}
    & 1+\frac{\Delta_\ell+\tilde{\Delta}_r}{2} \geq \lceil\tilde{i}^{-1}(\epsilon/2)\rceil = \Delta^\star(\epsilon/2)\\
    \Leftrightarrow& \Delta_\ell+\tilde{\Delta}_r \geq 2 \Delta^\star(\epsilon/2) - 1 .
\end{align*}
Finally, we note that
\begin{equation}
    \label{eq:rr_two-sided-leakage}
    \Delta_\ell+\Delta_r \geq \Delta_\ell+\tilde{\Delta}_r-1 \geq 2 \Delta^\star(\epsilon/2) - 2.
\end{equation}

The minimum number of redacted records is then bounded by the minimum of \cref{eq:rr_one-sided-leakage} and \cref{eq:rr_two-sided-leakage}. In particular, a two-sided release yields higher utility if
\begin{align*}
    (\ref{eq:rr_one-sided-leakage}) \geq (\ref{eq:rr_two-sided-leakage}) &\Leftrightarrow & \p+\Delta^\star(\epsilon)-2 &\geq 2 \Delta^\star(\epsilon/2) - 2
    \\
    &\Leftrightarrow & \p + \Delta^\star(\epsilon) - 2 \Delta^\star(\epsilon/2)
    &\geq 0.
\end{align*}
The threshold reveals the following fact: if a record is far away from the boundary of the Markov chain ($\p \leq n/2$ large) and the privacy budget is sufficiently large, two-sided leakage with a symmetric redaction radius is optimal. Otherwise, one-sided leakage is more efficient, and the shorter chain should be redacted completely.
Defining $t_\p(\epsilon) = \p + \Delta^\star(\epsilon) - 2 \Delta^\star(\epsilon/2)$, we can summarize Case 2 by the following bound:

\begin{align*}
    \Delta_\ell+\Delta_r &\geq
    \begin{cases}
        \p+\Delta^\star(\epsilon)-2, &\text{if } t_\p(\epsilon) < 0,
        \\
        2 \Delta^\star(\epsilon/2) - 2, &\text{if } t_\p(\epsilon) \geq 0,
    \end{cases}
    \\
    &= \min\left\{ \p+\Delta^\star(\epsilon), 2 \Delta^\star(\epsilon/2) \right\} - 2.
\end{align*}
Since $n (1-\nu_\mathrm{DIM}) \geq \Delta_\ell+\Delta_r+1$, this ultimately yields
\begin{equation*}
    \nu_\mathrm{DIM}
    \leq
    1 - \frac{1}{n} \left( \min\{
        2 \Delta^\star(\epsilon/2) -1,
        \p+\Delta^\star(\epsilon) -1
    \} \right).
\end{equation*}

\subsection{Lower bound on the privacy leakage}
In this section, we prove a lower bound on the privacy leakage of local redaction mechanisms.

\begin{lemma}[Leakage is lower-bounded by influence of released realizations]
\label{lma:zerohop_redaction}
Let $\mathcal{Q} \subseteq \range{n}$.
If $\x_{\mathcal{Q}} \in \supp(\Y_\mathcal{Q})$, then the privacy leakage about $X_\p$ of any local redaction mechanism is
\begin{equation*}
    \leak{X_\p \to \Y} \geq \pwInfl{X_\p \rightsquigarrow \X_{\mathcal{Q}}=\x_{\mathcal{Q}}}.
\end{equation*}
\end{lemma}

\begin{corollary}
\label{cor:leakage_bound}
For a \di local redaction mechanism, let $\mathcal{Q} \subseteq \region[rel]$. The leakage about $X_\p$ is
\begin{equation*}
    \leak{X_\p \to \Y} \geq \infl{X_\p \rightsquigarrow \X_{\mathcal{Q}}}.
\end{equation*}
If $\mathcal{Q}$ is a Markov quilt for $X_\p$, which has only redacted nearby records $\mathcal{N}$ with $\mathcal{N} \cap \region[rel] = \emptyset$ then equality holds.
\end{corollary}

The proof idea is to show that in the case where $\y$ with $\y_\mathcal{Q}=\x_\mathcal{Q}$ is a possible output, the leakage about $X_i$ is bounded from below by the min-influence from $X_i$ on $\X_\Q$. Thus, $y_t = \perp$ must hold for at least some $t \in \mathcal{Q}$.

Formally, let $\y \in \mathcal{Y}^n$ with $y_t \neq \perp$ for all $t \in \mathcal{Q}$ and suppose that $\y \in \supp(\Y)$. Define $\bar{\mathcal{Q}}=\range{n} \setminus (\mathcal{Q} \cup \{i\})$. We assume w.l.o.g. that $X_\p \notin \mathcal{Q}$ ($X_\p$ cannot be part of the revealed records since privacy of $X_\p$ always requires a redaction of $X_\p$). It holds:
\begin{align*}
    & \Pr{\Y=\y | X_\p=x}
    \\
    \overset{(a)}{=} & \Pr{\Y=\y | X_\p=x, \X_\mathcal{Q}=\y_\mathcal{Q}}
    \Pr{\X_\mathcal{Q}=\y_\mathcal{Q} | X_\p=x}
    \\
    \overset{(b)}{=} & \Pr{\Y_{\bar{\mathcal{Q}}}=\y_{\bar{\mathcal{Q}}} | X_\p=x, \X_\mathcal{Q}=\y_\mathcal{Q}}
    \\
    & \cdot \Pr{\X_\mathcal{Q}=\y_\mathcal{Q} | X_\p=x} \Pr{\Y_\mathcal{Q}=\y_\mathcal{Q} | \X_\mathcal{Q}=\y_\mathcal{Q}}
\end{align*}
where $(a)$ follows since $\Pr{Y_t=x_t | X_t=x}=0$ if $x \neq x_t$ for redaction mechanisms, and $(b)$ follows by the Markovity of the chain.
Hence, we can conclude that
\begin{align}
    \frac{\Pr{\Y=\y | X_\p=x}}{\Pr{\Y=\y | X_\p=\bar{x}}}
    =&\, \frac{
        \Pr{\Y_{\bar{\mathcal{Q}}}=\y_{\bar{\mathcal{Q}}} | X_\p=x, \X_\mathcal{Q}=\y_\mathcal{Q}}
    }{
        \Pr{\Y_{\bar{\mathcal{Q}}}=\y_{\bar{\mathcal{Q}}} | X_\p=\bar{x}, \X_\mathcal{Q}=\y_\mathcal{Q}}
    }
    \label{eq:term1}\\
    &\cdot \frac{ \Pr{\X_\mathcal{Q}=\y_\mathcal{Q} | X_\p=x} }{ \Pr{\X_\mathcal{Q}=\y_\mathcal{Q} | X_\p=\bar{x}} }.
    \label{eq:term2}
\end{align}

We choose
\begin{equation*}
    x = \arg \max_{x_\p \in \mathcal{X}} \frac{ \Pr{\X_\mathcal{Q}=\y_\mathcal{Q} | X_\p=x_\p} }{ \Pr{\X_\mathcal{Q}=\y_\mathcal{Q} | X_\p=\bar{x}_\p} },
\end{equation*}
and therefore, \cref{eq:term2} becomes
\begin{equation*}
    \frac{ \Pr{\X_\mathcal{Q}=\y_\mathcal{Q} | X_\p=x} }{ \Pr{\X_\mathcal{Q}=\y_\mathcal{Q} | X_\p=\bar{x}} }
    = \max_{x_\p \in \mathcal{X}} \frac{ \Pr{\X_\mathcal{Q}=\y_\mathcal{Q} | X_\p=x_\p} }{ \Pr{\X_\mathcal{Q}=\y_\mathcal{Q} | X_\p=\bar{x}_\p} }.
\end{equation*}
Note that $\max_{x_\p \in \mathcal{X}} \frac{ \Pr{\X_\mathcal{Q}=\y_\mathcal{Q} | X_\p=x_\p} }{ \Pr{\X_\mathcal{Q}=\y_\mathcal{Q} | X_\p=\bar{x}_\p} } < \infl{X_\p \rightsquigarrow \X_\mathcal{Q}}$ in general.
Finally, there always exists a $\y \in \supp(\Y)$ with values $\y_{\bar{\mathcal{Q}}}$ such that for \cref{eq:term1} it holds
\begin{equation*}
    \frac{
        \Pr{\Y_{\bar{\mathcal{Q}}}=\y_{\bar{\mathcal{Q}}} | X_\p=x, \X_\mathcal{Q}=\y_\mathcal{Q}}
    }{
        \Pr{\Y_{\bar{\mathcal{Q}}}=\y_{\bar{\mathcal{Q}}} | X_\p=\bar{x}, \X_\mathcal{Q}=\y_\mathcal{Q}}
    }
    \geq 1.
\end{equation*}
Otherwise, we have a contradiction:
\begin{align*}
    & \Pr{\Y_{\bar{\mathcal{Q}}}=\y_{\bar{\mathcal{Q}}} | X_\p=x, \X_\mathcal{Q}=\y_\mathcal{Q}}
    \\
    &< \Pr{\Y_{\bar{\mathcal{Q}}}=\y_{\bar{\mathcal{Q}}} | X_\p=\bar{x}, \X_\mathcal{Q}=\y_\mathcal{Q}} \forall_{\y_{\bar{\mathcal{Q}}}}
    \\
    \implies& \sum_{\y_{\bar{\mathcal{Q}}}} \Pr{\Y_{\bar{\mathcal{Q}}}=\y_{\bar{\mathcal{Q}}} | X_\p=x, \X_\mathcal{Q}=\y_\mathcal{Q}}
    \\
    &< 
    \sum_{\y_{\bar{\mathcal{Q}}}} \Pr{\Y_{\bar{\mathcal{Q}}}=\y_{\bar{\mathcal{Q}}} | X_\p=\bar{x}, \X_\mathcal{Q}=\y_\mathcal{Q}} = 1.
\end{align*}
In summary, we can conclude that
\begin{equation*}
    \leak{X_\p \to \Y} \geq \min_{\y_\mathcal{Q} \in \mathcal{X}^{|\mathcal{Q}|}} \max_{x_\p \in \mathcal{X}} \frac{ \Pr{\X_\mathcal{Q}=\y_\mathcal{Q} | X_\p=x_\p} }{ \Pr{\X_\mathcal{Q}=\y_\mathcal{Q} | X_\p=\bar{x}_\p} }.
\end{equation*}

For \di local redaction mechanisms, $\x_\mathcal{Q} \in \supp(\Y_\mathcal{Q})$ for all $\x_\mathcal{Q} \in \mathcal{X}^{|\mathcal{Q}|}$ whenever $\mathcal{Q} \subseteq \region[rel]$. Thus, the lower bound follows from \cref{lma:zerohop_redaction} by choosing
$\x_\mathcal{Q} = \arg \max_{\x_\mathcal{Q}^\prime \in \mathcal{X}^{|\mathcal{Q}|}} \pwInfl{X_\p \rightsquigarrow \X_{\mathcal{Q}}=\x_{\mathcal{Q}}^\prime}$. Furthermore, $\leak{X_\p \to \Y} \leq \infl{X_\p \rightsquigarrow \X_{\mathcal{Q}}}$ follows by the data processing inequality of \ac{ldp} since $\Y$ depends on $X_\p$ only through $\X_\mathcal{Q}$.

\subsection{Properties of the pointwise-influence and max-influence}
\label{apx:influence_properties}

\begin{proposition}[Influence based on distance]
    \label{prop:max_influence_by_distance}
    Let $\p,t \in \range{n}$ and $|\p-t|=\Delta$ for $\Delta>0$. Then it holds that
    \begin{itemize}
        \item $\pwInfl{X_\p \rightsquigarrow X_t=0} = \lowInfl{\Delta}$,
        \item $\pwInfl{X_\p \rightsquigarrow X_t=1} = \highInfl{\Delta}$,
        \item $\infl{X_\p \rightsquigarrow X_t} = \highInfl{\Delta}$,
    \end{itemize}
    with
    \begin{align}
        \label{eq:infl_low}
        \lowInfl{\Delta} &\define \left| \log \frac{1+ \frac{\alpha}{\beta} (1-\alpha-\beta)^\Delta}{1- (1-\alpha-\beta)^\Delta} \right|,
        \\
        \label{eq:infl_high}
        \highInfl{\Delta} &\define \left| \log \frac{1+ \frac{\beta}{\alpha} (1-\alpha-\beta)^\Delta}{1- (1-\alpha-\beta)^\Delta} \right|.
    \end{align}
    The sequences $\highInfl{\Delta}$ and $\lowInfl{\Delta}$ with $\Delta \in \mathbb{N}$ are convex, i.e., their affine extensions are convex.
\end{proposition}

We prove the statements in \cref{prop:max_influence_by_distance} in the sequel.
By the stationarity of the Markov chain, the influence only depends on the difference $\Delta=|\p-t|$. We consider only the case $t>\p$ here. As $\Pr{X_{t+1}=j | X_t=i} = \Pr{X_t=j | X_{t+1}=i}$ holds by the stationary distribution assumption, the case $t<\p$ follows along the same line of arguments and yields the same result.
The likelihood ratios can be derived from the transition matrix $P$ of the Markov chain:
\begin{align*}
    \frac{
        \Pr{ X_t = j | X_\p = i}
    }{
        \Pr{ X_t = j | X_\p = k}
    }
    =&
    \frac{
        \Pr{ X_{\p+\Delta} = j | X_\p = i}
    }{
        \Pr{ X_{\p+\Delta} = j | X_\p = k}
    }
    =
    \frac{
        (P^\Delta)_{ij}
    }{
        (P^\Delta)_{kj}
    },
\end{align*}
where we can compute
\begin{equation*}
    P^\Delta
    =
    \begin{pmatrix}
        1-\alpha & \alpha \\
        \beta & 1-\beta
    \end{pmatrix}^\Delta
    =
    \begin{pmatrix}
        1-\alpha_\Delta & \alpha_\Delta \\
        \beta_\Delta & 1-\beta_\Delta
    \end{pmatrix}
\end{equation*}
with $\alpha_\Delta \define \frac{\alpha}{\alpha+\beta} (1-(1-(\alpha+\beta))^\Delta)$
and $\beta_\Delta \define \frac{\beta}{\alpha+\beta} (1-(1-(\alpha+\beta))^\Delta)$. Observe from the definition of $\alpha_\Delta$ and $\beta_\Delta$ that
\begin{align}
    \label{eq:alpha_beta_ratio_const}
    \alpha_\Delta/\beta_\Delta &= \alpha/\beta,
    \\
    1-\alpha_\Delta-\beta_\Delta &= (1-\alpha-\beta)^\Delta.
\end{align}

\paragraph*{Closed-form expressions}
The closed-form expressions can be calculated straightforwardly as

\begin{align*}
    \nonumber
    &\pwInfl{X_\p \rightsquigarrow X_t=0} 
    = \log \max_{x \in \mathcal{X}} \frac{ \Pr{X_t=0 | X_i=x} }{ \Pr{X_t=0 | X_i=\bar{x}} }
    \\
    =& \log \max\left\{ \frac{1-\alpha_\Delta}{\beta_\Delta}, \frac{\beta_\Delta}{1-\alpha_\Delta} \right\}
    = \left| \log \frac{1-\alpha_\Delta}{\beta_\Delta} \right|
    \\
    =& \left| \log \frac{1+\frac{\alpha}{\beta} (1-\alpha-\beta)^\Delta}{1- (1-\alpha-\beta)^\Delta} \right|,
\end{align*}
for $x_t=0$. The analog derivation applies for $x_t=1$.
It can be easily verified that $\pwInfl{X_\p \rightsquigarrow X_t=0} \leq \pwInfl{X_\p \rightsquigarrow X_t=1}$ if $\alpha \leq \beta$.

\paragraph*{Convexity}
To prove the convexity, we consider the generalizing function
\begin{equation}
\label{eq:unified_i}
i_c(\Delta)=\left| \log \frac{1+c (1-\alpha-\beta)^\Delta}{1- (1-\alpha-\beta)^\Delta} \right|
\end{equation}
with $\lowInfl{\Delta} = i_{\alpha/\beta}(\Delta)$ and $\highInfl{\Delta} = i_{\beta/\alpha}(\Delta)$.

For $1-\alpha-\beta \geq 0$, we can show that $i_c(\Delta)$ with $c \in \{\frac{\alpha}{\beta},\frac{\beta}{\alpha}\}$ is convex in $\Delta \in \mathbb{R}_{>0}$ by showing that the second derivative is positive:
\begin{align*}
    \hat{i}^{\prime\prime}(\Delta) = \left( \log(1-\alpha-\beta) \right)^2 (1-\alpha-\beta)^\Delta \left[ A+B \right]\\
    A = \frac{c}{\left(1+c (1-\alpha-\beta)^\Delta\right)^2},\,
    B = \frac{1}{\left(1- (1-\alpha-\beta)^\Delta\right)^2}.
\end{align*}
The second derivative is positive since $(1-\alpha-\beta)^\Delta \geq 0$.

For $1-\alpha-\beta < 0$, we show that $i(\Delta)=i_c(\Delta)-i_c(\Delta+1)$ with $c \in \{\frac{\alpha}{\beta},\frac{\beta}{\alpha}\}$ is decreasing in $\Delta$.

\underline{$\Delta$ even:} In this case, $\Delta+1$ is odd, and we can write
\begin{align}
    i(\Delta)
    &= \log\frac{1+c(1-\alpha-\beta)^\Delta}{1-(1-\alpha-\beta)^\Delta} - \log\frac{1-(1-\alpha-\beta)^{\Delta+1}}{1+c(1-\alpha-\beta)^{\Delta+1}}
    \nonumber
    \\
    &= \log \frac{1+c|1-\alpha-\beta|^\Delta}{1+d|1-\alpha-\beta|^\Delta} + \log \frac{1-cd|1-\alpha-\beta|^\Delta}{1-|1-\alpha-\beta|^\Delta},
    \label{eq:even_condition}
\end{align}
with $d = |1-\alpha-\beta| > 0$.

\underline{$\Delta$ odd:} In this case, $\Delta+1$ is even, and we can apply the same steps and obtain
\begin{align}
    i(\Delta)
    &= \log \frac{1+|1-\alpha-\beta|^\Delta}{1+cd|1-\alpha-\beta|^\Delta} + \log \frac{1-d|1-\alpha-\beta|^\Delta}{1-c|1-\alpha-\beta|^\Delta}.
    \label{eq:odd_condition}
\end{align}
In both cases, the argument of the logarithm is positive since $c,d > 0$, $0 < |1-\alpha-\beta| < 1$, and $cd < 1$.
While the first two conditions hold by definition, the last condition is equivalent to
\begin{align*}
    cd < 1 \Leftrightarrow \frac{\alpha}{\beta}|1-\alpha-\beta| \leq \frac{\beta}{\alpha}|1-\alpha-\beta| < 1.
\end{align*}
This inequality holds since $\alpha \leq \beta$ and since for $1-\alpha-\beta < 0$:
\begin{align*}
    &\frac{\beta}{\alpha} |1-\alpha-\beta|
    = \frac{\beta}{\alpha} (\alpha+\beta-1)
    = \beta - \frac{\beta}{\alpha} (1-\beta)
    \\
    =& 1-(1-\beta) - \frac{\beta}{\alpha} (1-\beta)
    = 1 - \frac{\alpha+\beta}{\alpha} (1-\beta) < 1.
\end{align*}

The function $i(\Delta)$ is composed of functions of the class
\begin{equation*}
    f_{a,b}(\Delta) = \frac{1+a \cdot |1-\alpha-\beta|^\Delta}{1+b \cdot |1-\alpha-\beta|^\Delta},\,a,b \in \mathbb{R}.
\end{equation*}
The derivative is
\begin{equation*}
    f_{a,b}^\prime(\Delta) = \frac{(a-b) \log|1-\alpha-\beta| \cdot |1-\alpha-\beta|^\Delta}{(1+b \cdot |1-\alpha-\beta|^\Delta)^2},
\end{equation*}
which is negative if $a > b$ since $0 < |1-\alpha-\beta| < 1$ and $\log|1-\alpha-\beta| < 0$. Therefore, we can conclude that $i(\Delta)$ is decreasing in $\Delta$ if $a > b$ in \cref{eq:even_condition} and \cref{eq:odd_condition}, respectively. In particular, this requires $cd < 1$ and $c > d$. One can verify that these statements are equivalent for $\alpha \leq \beta$, and the former has already been verified before.


\begin{thebibliography}{10}

\bibitem{songPufferfishPrivacyMechanisms2017a}
S.~Song, Y.~Wang, and K.~Chaudhuri, ``Pufferfish {{Privacy Mechanisms}} for
  {{Correlated Data}},'' in {\em Proceedings of the 2017 {{ACM International
  Conference}} on {{Management}} of {{Data}}}, pp.~1291--1306, ACM, May 2017.

\bibitem{europeanparliamentRegulationEU2016}
{European Parliament} and {Council of the European Union}, ``Regulation
  ({{EU}}) 2016/679 of the {{European Parliament}} and of the {{Council}}. of
  27 {{April}} 2016 on the protection of natural persons with regard to the
  processing of personal data and on the free movement of such data, and
  repealing {{Directive}} 95/46/{{EC}} ({{General Data Protection
  Regulation}}).'' \url{https://data.europa.eu/eli/reg/2016/679/oj}, May 2016.

\bibitem{ccpa2018}
{State of California Department of Justice}, ``California consumer privacy act
  of 2018.''
  \url{https://leginfo.legislature.ca.gov/faces/codes_displayText.xhtml?lawCode=CIV&division=3.&title=1.81.5.&part=4.},
  2018.

\bibitem{liuLocationPrivacyIts2018}
B.~Liu, W.~Zhou, T.~Zhu, L.~Gao, and Y.~Xiang, ``Location {{Privacy}} and {{Its
  Applications}}: {{A Systematic Study}},'' {\em IEEE Access}, vol.~6,
  pp.~17606--17624, 2018.

\bibitem{lucaHowSocialRelationships2010}
E.~W. De~Luca, A.~Said, and S.~Albayrak, ``How social relationships affect user
  similarities,'' in {\em Proceedings of the {{Social Recommender Systems
  Workshop}} ({{SRS2010}}), in Conjunction with the 2010 {{International
  Conference}} on {{Intelligent User Interfaces}}.}, 2010.

\bibitem{bonhardKnowingMeKnowing2006}
P.~Bonhard and M.~A. Sasse, ``'{{Knowing}} me, knowing you' --- {{Using}}
  profiles and social networking to improve recommender systems,'' {\em BT
  Technology Journal}, vol.~24, pp.~84--98, July 2006.

\bibitem{rebeccacarballoDataBreach23andMe2023}
R.~Carballo, ``Data breach at {{23andMe}} affects 6.9 million profiles, company
  says.'' \url{https://www.nytimes.com/2023/12/04/us/23andme-hack-data.html},
  Dec. 2023.

\bibitem{yeMechanismsHidingSensitive2022}
F.~Ye, H.~Cho, and S.~El~Rouayheb, ``Mechanisms for {{Hiding Sensitive
  Genotypes With Information-Theoretic Privacy}},'' {\em IEEE Transactions on
  Information Theory}, vol.~68, pp.~4090--4105, June 2022.

\bibitem{kiferNoFreeLunch2011}
D.~Kifer and A.~Machanavajjhala, ``No free lunch in data privacy,'' in {\em
  Proceedings of the 2011 {{ACM SIGMOD International Conference}} on
  {{Management}} of Data}, (New York, NY, USA), pp.~193--204, ACM, June 2011.

\bibitem{chenCorrelatedNetworkData2014}
R.~Chen, B.~C.~M. Fung, P.~S. Yu, and B.~C. Desai, ``Correlated network data
  publication via differential privacy,'' {\em The VLDB Journal}, vol.~23,
  pp.~653--676, Aug. 2014.

\bibitem{xiaoProtectingLocationsDifferential2015}
Y.~Xiao and L.~Xiong, ``Protecting {{Locations}} with {{Differential Privacy}}
  under {{Temporal Correlations}},'' in {\em Proceedings of the 22nd {{ACM
  SIGSAC Conference}} on {{Computer}} and {{Communications Security}}}, (New
  York, NY, USA), pp.~1298--1309, ACM, Oct. 2015.

\bibitem{yangBayesianDifferentialPrivacy2015}
B.~Yang, I.~Sato, and H.~Nakagawa, ``Bayesian {{Differential Privacy}} on
  {{Correlated Data}},'' in {\em Proceedings of the 2015 {{ACM SIGMOD
  International Conference}} on {{Management}} of {{Data}}}, (New York, NY,
  USA), pp.~747--762, ACM, May 2015.

\bibitem{zhuCorrelatedDifferentialPrivacy2015}
T.~Zhu, P.~Xiong, G.~Li, and W.~Zhou, ``Correlated {{Differential Privacy}}:
  {{Hiding Information}} in {{Non-IID Data Set}},'' {\em IEEE Transactions on
  Information Forensics and Security}, vol.~10, pp.~229--242, Feb. 2015.

\bibitem{liuDependenceMakesYou2016}
C.~Liu, S.~Chakraborty, and P.~Mittal, ``Dependence {{Makes You Vulnerable}}:
  {{Differential Privacy Under Dependent Tuples}},'' in {\em Proceedings of the
  2016 {{Network}} and {{Distributed System Security Symposium}}}, (San Diego,
  CA), Internet Society, 2016.

\bibitem{ghoshInferentialPrivacyGuarantees2017}
A.~Ghosh and R.~Kleinberg, ``Inferential privacy guarantees for differentially
  private mechanisms,'' in {\em Proceedings of the 8th Innovations in
  Theoretical Computer Science Conference} (C.~H. Papadimitriou, ed.), vol.~67
  of {\em Leibniz International Proceedings in Informatics (Lipics)},
  (Dagstuhl, Germany), pp.~9:1--9:3, Schloss Dagstuhl--Leibniz-Zentrum fuer
  Informatik, 2017.

\bibitem{chakrabartiOptimalLocalBayesian2022}
D.~Chakrabarti, J.~Gao, A.~Saraf, G.~Schoenebeck, and F.-Y. Yu, ``Optimal
  {{Local Bayesian Differential Privacy}} over {{Markov Chains}},'' {\em arXiv
  preprint}, June 2022.

\bibitem{dworkDifferentialPrivacy2006}
C.~Dwork, ``Differential {{Privacy}},'' in {\em Proceedings of the
  International Colloquium on Automata, Languages, and Programming}
  (M.~Bugliesi, B.~Preneel, V.~Sassone, and I.~Wegener, eds.), Lecture
  {{Notes}} in {{Computer Science}}, (Berlin, Heidelberg), pp.~1--12, Springer,
  2006.

\bibitem{zhaoDependentDifferentialPrivacy2017}
J.~Zhao, J.~Zhang, and H.~V. Poor, ``Dependent {{Differential Privacy}} for
  {{Correlated Data}},'' in {\em Proceedings of the 2017 {{IEEE Globecom
  Workshops}} ({{GC Wkshps}})}, pp.~1--7, Dec. 2017.

\bibitem{naimONOFFPrivacyCorrelated2019a}
C.~Naim, F.~Ye, and S.~E. Rouayheb, ``{{ON-OFF Privacy}} with {{Correlated
  Requests}},'' in {\em Proceedings of the 2019 {{IEEE International
  Symposium}} on {{Information Theory}}}, pp.~817--821, July 2019.

\bibitem{yeIntermittentPrivateInformation2022}
F.~Ye and S.~El~Rouayheb, ``Intermittent {{Private Information Retrieval With
  Application}} to {{Location Privacy}},'' {\em IEEE Journal on Selected Areas
  in Communications}, vol.~40, pp.~927--939, Mar. 2022.

\bibitem{jiangAnsweringCountQueries2023}
B.~Jiang, M.~Seif, R.~Tandon, and M.~Li, ``Answering {{Count Queries}} for
  {{Genomic Data With Perfect Privacy}},'' {\em IEEE Transactions on
  Information Forensics and Security}, vol.~18, pp.~3862--3875, 2023.

\bibitem{naimPrivacySocialNetworks2023}
C.~Naim, F.~Ye, and S.~El~Rouayheb, ``On the {{Privacy}} of {{Social Networks}}
  with {{Personal Privacy Choices}},'' in {\em Proceedings of the 2023 {{IEEE
  International Symposium}} on {{Information Theory}}}, pp.~1812--1817, June
  2023.

\bibitem{duchiLocalPrivacyStatistical2013}
J.~C. Duchi, M.~I. Jordan, and M.~J. Wainwright, ``Local {{Privacy}} and
  {{Statistical Minimax Rates}},'' in {\em Proceedings of the 2013 {{IEEE}}
  Symposium on {{Foundations}} of {{Computer Science}}}, pp.~429--438, Oct.
  2013.

\bibitem{kiferPufferfishFrameworkMathematical2014}
D.~Kifer and A.~Machanavajjhala, ``Pufferfish: {{A}} framework for mathematical
  privacy definitions,'' {\em ACM Transactions on Database Systems}, vol.~39,
  pp.~3:1--3:36, Jan. 2014.

\bibitem{blochOverviewInformationTheoreticSecurity2021}
M.~Bloch, O.~G{\"u}nl{\"u}, A.~Yener, F.~Oggier, H.~V. Poor, L.~Sankar, and
  R.~F. Schaefer, ``An {{Overview}} of {{Information-Theoretic Security}} and
  {{Privacy}}: {{Metrics}}, {{Limits}} and {{Applications}},'' {\em IEEE
  Journal on Selected Areas in Information Theory}, vol.~2, pp.~5--22, Mar.
  2021.

\bibitem{wagnerTechnicalPrivacyMetrics2018}
I.~Wagner and D.~Eckhoff, ``Technical {{Privacy Metrics}}: {{A Systematic
  Survey}},'' {\em ACM Computing Surveys}, vol.~51, pp.~57:1--57:38, June 2018.

\bibitem{boyd2004convex}
S.~Boyd and L.~Vandenberghe, {\em Convex optimization}.
\newblock Cambridge University Press, 2004.

\end{thebibliography}
\end{document}